\documentclass[aps,pra,letterpaper,superscriptaddress,twocolumn,floatfix]{revtex4-2}
\usepackage{graphicx}
\usepackage{dcolumn}
\usepackage{url}
\usepackage{bm}
\usepackage{makecell}
\usepackage{graphicx} 
\usepackage{tabularx} 
\usepackage{subfigure}
\usepackage{xcolor}
\newcommand{\be}{\begin{eqnarray}}
\newcommand{\ee}{\end{eqnarray}}
\usepackage{amsfonts}
\usepackage{bbm}
\usepackage{amsmath,leftidx}
\usepackage{graphicx}
\usepackage{times}
\usepackage{CJK}
\usepackage{color,colortbl}
\usepackage[normalem]{ulem}
\usepackage{hyperref}
\usepackage{soul}
\setstcolor{red}

\graphicspath{./}

\begin{document}
\title{
Variational Tensor Network Operator 
}

\author{Yu-Hsueh Chen}
\affiliation{Center for Theoretical Physics and Department of Physics, National Taiwan University, Taipei, 10607, Taiwan}
\author{Ke Hsu}
\affiliation{Center for Theoretical Physics and Department of Physics, National Taiwan University, Taipei, 10607, Taiwan}
\author{Wei-Lin Tu}
\affiliation{Division of Display and Semiconductor Physics, Korea University, Sejong 30019, Korea}
\author{Hyun-Yong Lee}
\affiliation{Division of Display and Semiconductor Physics, Korea University, Sejong 30019, Korea}
\affiliation{Department of Applied Physics, Graduate School, Korea University, Sejong 30019, Korea}
\affiliation{Interdisciplinary Program in E·ICT-Culture-Sports Convergence, Korea University, Sejong 30019, Korea}
\author{Ying-Jer Kao}
\affiliation{Center for Theoretical Physics and Department of Physics, National Taiwan University, Taipei, 10607, Taiwan}
\affiliation{Center for Quantum Science and Technology,  National Taiwan University, Taipei, 10607, Taiwan}
\begin{abstract}
We propose a simple and generic construction of the variational tensor network operators to study the quantum  spin systems by the synergy of ideas from the imaginary-time evolution and variational optimization of trial wave functions.
 By applying these operators to  simple initial states, accurate variational ground state wave functions with extremely few parameters can be obtained. 
Furthermore, the framework can be applied to study spontaneously symmetry breaking, symmetry protected topological, and intrinsic topologically ordered phases, and we show that symmetries of the local tensors associated with these phases can emerge directly after the optimization without any gauge fixing. 
This provides a universal way to identify quantum phase transitions without prior knowledge of the system.

\end{abstract}
\maketitle

\section{Introduction}

Tensor networks (TN) are preeminent theoretical and computational approaches to study quantum many-body systems \citep{Orus_2019_review}. 
Among them, matrix product states (MPS) and projected entangled pair states (PEPS) receive particular attention to describe the ground states of  many-body systems in the thermodynamic limit {in one and higher dimensions} \citep{Review_TN_2021, Orus_2014_review}. 
The success of MPS and PEPS lies in the fact that they efficiently parameterize the area-law states and encapsulate all the information in a single tensor. 
The former guarantees that they are successful ansatz to represent the ground states of gapped local Hamiltonians, and the latter reduces the problem of classifying phases of matter into studying the symmetries of the local tensors. 
However, in practical simulations, significant efforts are still devoted to  address the following two questions: {\textit{Given the translationally invariant Hamiltonian, how to optimize the MPS and PEPS ansatz? {And after} the optimization, how to classify different phases in terms of the local tensors? }}

One natural way to search for the optimal ground states is to variationally optimize  all the parameters of the local tensor with proper gauge-fixing conditions. 
This includes the well-known density matrix renormalization group (DMRG) and the variational uniform matrix product state (VUMPS) {for MPS} \citep{White_1992_DMRG, White_1993_DMRG,2008_Ian_iDMRG,Zauner_2018_VUMPS}, as well as the gradient-based optimization {for PEPS}, where the gradients of the parameters are computed using either the sum of several tensor diagrams or the differentiable programming techniques \citep{Corboz_2016_Var, Vanderstraeten_2016_GD, Liao_2019_Diff}. 
While variational optimization (VarOpt) makes full use of the TN ansatz and yields the most reliable ground states, it is computationally demanding, and the optimization process may potentially be trapped to local minima due to a large number of free parameters. 
This downside is particularly severe for the 2D PEPS and the higher-dimensional tensor networks.
Another method to search for the optimal tensor is to perform the imaginary time evolution (ITE). 
The basic idea is that the ground state can be obtained by evolving a random initial state with the operator $e^{-\tau \hat H}$ for a sufficiently long time. 
While the resulting state can be represented as a TN with infinite bond dimensions using the Trotter decomposition, the area law guarantees that one can reasonably truncate the bond dimensions. 
This method contains the infinite time-evolving block decimation (iTEBD) in 1D \citep{Vidal_2003_ITE, Vidal_2004_ITE, Vidal_2007_ITE} and the simple and full updates method in 2D \citep{Jiang_2008_ITE, Jordan_2008_ITE, Zhao_2010_fullupdate, Phien_2015_fullupdate}. 
ITE method is computationally cheap as it contains no variational parameters. 
However,  the Trotter error due to the discretization in time and the necessary truncation  leads to loss of accuracy. 
Therefore, the TN wave functions obtained from ITE are usually biased by the choice of initial state and may not capture the long-range correlation and entanglement accurately.
%

After the optimization, a recurring theme is to determine the phase of the system from the optimized tensor. 
While the classification of several quantum phases in terms of MPS and PEPS are well developed \citep{Schuch_PG_2011,Chen_2011_1Dclassification,Schuch_2010_Ginjective,Williamson_2016_SPT,Bultinck_2017_anyonMPO}, implementing those theories in practice requires prior knowledge of the system. 
Furthermore, the techniques developed for a particular phase of matter cannot be easily generalized to  others. 
This {comes from} the fact that the MPS and PEPS representations are not unique, and the characteristics of their building block tensors in different phases become clearer in different gauge choices.
For example, since the one-dimensional symmetry-protected topologically ordered (SPT) phases are characterized by the projective representations of the symmetry group, the specific numerical approach was developed to extract the corresponding representations from the MPS \citep{Pollmann_2012_SPTdetection}. 
On the other hand, the intrinsic topologically ordered phases (TO) are characterized by the global symmetries on the virtual bonds, and one should impose the virtual symmetry during the optimization \citep{He_2014_GSPTRG,Iqbal_2021_orderparam}. 
We note that some unbiased techniques to study the topologically ordered phases without imposing symmetries during the optimization were proposed recently \citep{Crone_2020_toriccode, Francuz_2020_determineAbelian}. 
However, one should perform tedious gauge fixing procedures after the optimization, and this still requires the knowledge {of what kind of TO phase we are studying}.

In this paper, we address the aforementioned questions by proposing a simple and generic construction of the variational tensor network operator (TNO) which can be derived merely from the knowledge of the model Hamiltonian. 
We refer to this operator as Generic-TNO or GTNO in the following.
The {generic} tensor network state (GTNS) containing variational parameters is then obtained by applying GTNO to some simple initial states.
GTNO combines the idea of ITE and VarOpt by constructing a variational operator which projects the simple initial state to the ground state.
Specifically, consider the translationally invariant Hamiltonian of the form $\hat{H} = \sum h^{(l)}_{i_1,i_2,...,i_l}$, where $h^{(l)}_{i_1,i_2,...,i_l}$ is the local interaction with an $l$-site support.
The method of ITE, i.e., applying $e^{-\tau \hat H}$ for a sufficiently long time, implies that the ground state can be obtained by applying some linear combinations of the operators generated from the Hamiltonian. 
While in general one cannot construct the exact ground-state projector efficiently from $\hat{H}$,
one can use the local interaction $h^{(l)}_{i_1,i_2,...,i_l}$ to build a variational operator satisfying all the symmetries of $\hat H$.
The key insight of GTNO is that all the powers of $h^{(l)}_{i_1,i_2,...,i_l}$ can be decomposed into the smaller tensors. 
It is then natural to construct a variational TNO using those tensors as the building blocks.
On the side of obtaining ground states, GTNS needs far fewer parameters but yields comparable energy with VarOpt. 
On the side of classifying phases of matter, symmetries of the local GTNS emerge themselves after the optimization. 
%
{Different from most of the other numerical approaches that actively look for the symmetries of the local tensor, GTNS let those symmetries come looking for it after the optimization, which is extraordinary.}
As we will demonstrate below, without strictly imposing any symmetry before or performing gauge fixing after the optimization, GTNS can be used to distinguish several quantum phases, including the spontaneously symmetry breaking (SSB), SPT, and the intrinsic TO phases. 
Our study provides a unified protocol to study the general quantum spin systems and paves the way to explore new exotic phases. 

{The paper is organized as follows. In Sec.\ref{sec:central_idea}, we motivate the construction of GTNO by noting that one can exponetiate the local interaction $h^{(l)}_{i_1,i_2,...,i_l}$ to build  an efficient operator to project the initial state to the ground state. 
We then show that any power of the two-site interaction can be expressed as the contraction of two rank-3 tensors, which form the building blocks of the GTNO.
In Sec.\ref{sec:GMPO}, we explicitly construct the 1D GTNO, the generic matrix product operator (GMPO) from those building block tensors and discuss the symmetry properties  {as well as} the physical meaning.
After establishing the GMPO form, we demonstrate its power by studying the 1D Heisenberg, transverse field Ising (TFI), and transverse field cluster (TFC) models.
In Sec.\ref{sec:GPEPO}, we generalize the GTNO to the 2D systems, the generic projected entangled product operator (GPEPO).
The GPEPO is then applied to study the 2D Heisenberg, TFI, and the toric code model in a magnetic field.
There, we also compare the results obtained by the GPEPO framework with other numerical approaches.
Finally, we conclude in Sec.\ref{sec:discussion} and discuss several possible extensions of the present work.
}

\section{Motivation}
\label{sec:central_idea}
To motivate the construction of GTNO using the local interaction $h^{(l)}_{i_1,i_2,...,i_l}$, we first consider the frustration-free Hamiltonian, i.e., $[h^{(l)}_{i_1,i_2,...,i_l}, h^{(l)}_{i'_1,i'_2,...,i'_l}] = 0,\ \forall i,i'$, for simplicity.
The exact ground state projector $ \lim_{\tau \rightarrow \infty} e^{-\tau \hat{H} }$ can then be decomposed into the tensor product of all the local projectors $\lim_{\tau \rightarrow\infty} e^{-\tau (h^{(l)}_{i_1,i_2,...,i_l})}$, which can be further expanded as 
\begin{equation}
\label{eqn:expand_h}
\lim_{\tau \rightarrow\infty} e^{-\tau (h^{(l)}_{i_1,i_2,...,i_l})} 
=
\lim_{\tau \rightarrow\infty} \sum_{\alpha = 0}^\infty \frac{(-\tau)^\alpha}{\alpha!} (h^{(l)}_{i_1,i_2,...,i_l})^\alpha.
\end{equation}
Therefore, the exact projector to the ground state subspace can be written as the sum of tensor products of $h^{(l)}_{i_1,i_2,...,i_l}$ with proper coefficients. 
On the other hand, if the Hamiltonian is not frustration-free, the exact ground state projector should be written as the power series expansion of the global Hamiltonian $\hat{H}$:
\begin{equation}
\label{eqn:expand_expH}
 \lim_{\tau \rightarrow \infty} \sum_{m = 0}^{\infty} \frac{(-\tau)^m}{m!} \hat{H}^m = \lim_{\tau \rightarrow \infty} \sum_{m = 0}^{\infty} \frac{(-\tau)^m}{m!} \big( \sum_i h^{(l)}_{i_1,i_2,...,i_l}\big)^m.
\end{equation}
Since $[h^{(l)}_{i_1,i_2,...,i_l}, h^{(l)}_{i'_1,i'_2,...,i'_l}] \neq 0$ in general, $\hat{H}^m$  produces some operators that cannot be written as a tensor product of local terms, inhibiting an efficient TN representation.
However, the local nature of $h^{(l)}_{i_1,i_2,...,i_l}$ implies that most of the terms expanded in Eq.~(\ref{eqn:expand_expH}) do have a tensor-product structure.
Therefore, it is still possible to establish an efficient TNO using merely the power of $h^{(l)}_{i_1,i_2,...,i_l}$ rather than $\hat{H}$. 
As we will show in Sec.\ref{sec:GMPO} and \ref{sec:GPEPO}, the non-commuting property of the local interaction can be remedied using a symmetrizing procedure in the GTNO framework.

%
%
%
%
The key observation of our work is that all the powers of $h^{(l)}_{i_1,i_2,...,i_l}$ can be decomposed into the smaller tensors. 
The variational operator can then be naturally written as a TNO composed of those tensors.
In the following, we demonstrate how to obtain those building block tensors by considering the $l = 2$ case, i.e., the nearest-neighbor interactions, and show that they naturally inherit the on-site symmetries from the Hamiltonian.

\begin{figure}
 \centering
\includegraphics[width=\linewidth]{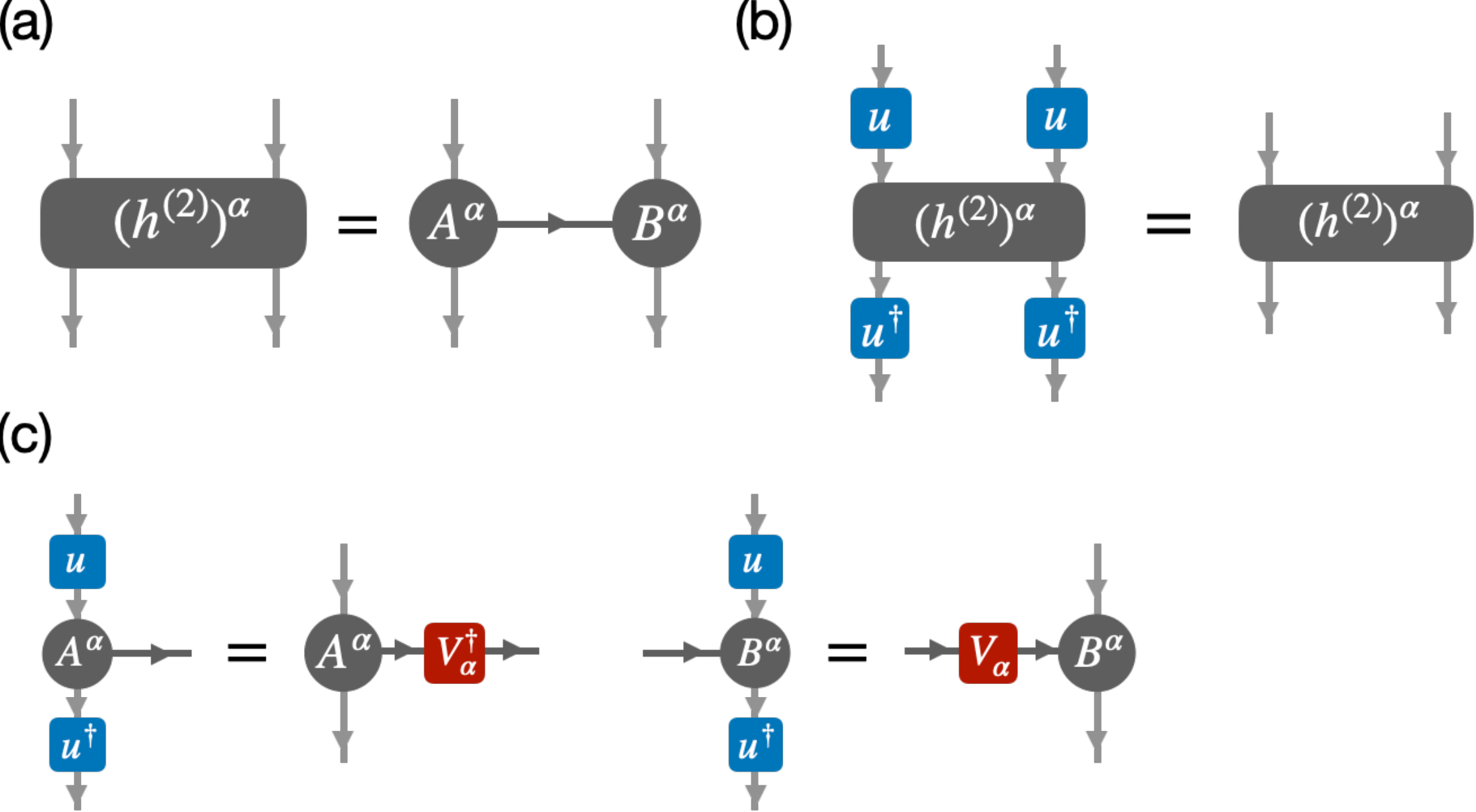}
\caption{(a) The $\alpha$-th power of the two-site interactions $h^{(2)}$ can always be decomposed into the contraction of two three-leg tensors $(A^\alpha|$ and $|B^\alpha)$. (b) If the Hamiltonian $\hat{H}$ respects the global on-site symmetry $\hat{U} = \prod_i u_i$, then $(h^{(2)}_{i,j})^\alpha$ is invariant under the local on-site action. (c) Under the physical transformation, $(A^\alpha|$ and $|B^\alpha)$ transforms like the bra and ket vectors in the virtual Hilbert space, respectively.} 
\label{fig:ph2_A_B}
\end{figure}

%
For any two-site interaction $h_{i,j}^{(2)}$, one can express the $\alpha$-th power of it as a sum of the tensor product of two local operators:
\begin{equation}
 (h^{(2)}_{i,j})^\alpha  = \sum_{v = 0}^{D_\alpha-1}   (A^\alpha_{v})_i \otimes (B^\alpha_{v})_j = (A^\alpha|B^\alpha),
\end{equation}
where $A^\alpha_v = \sum_{s_i s'_i} (A^\alpha)^{s_is'_i}_v |s_i\rangle \langle s'_i| $ is a physical operator acting on the site $i$ (similar relation for $B^\alpha_v$) [Fig.~\ref{fig:ph2_A_B}(a)]. 
In the second equality, we regard the collection of $(A^\alpha_0,A^\alpha_1,\cdots,A^\alpha_{D_\alpha-1})$ and $(B^\alpha_0,B^\alpha_1,\cdots,B^\alpha_{D_\alpha-1})$ as the \textit{vector operators} $(A^\alpha|$ and $|B^\alpha)$,  i.e., the vector in the virtual space $v$ with its element as the physical operator. 
Here, we use $|\cdot)$ to denote a  vector in the virtual Hilbert space as opposed to the physical one $|\cdot \rangle$.
It then follows that $(A^\alpha |B^\alpha)$ denotes the inner product in the virtual space.

If the Hamiltonian respects the global on-site symmetry $\mathbb{G}$: $\hat U(g) \hat H \hat U^\dagger(g) = \hat H,\ \forall g$ in $G$ with $\hat U(g) = \prod_{i} u_i(g)$, where $u_i(g)$ is the local representation of $g$, then $(h^{(2)}_{i,j})^\alpha$ is invariant under the local on-site transformation:
\begin{equation}
[u_i(g) \otimes u_j(g)] (h^{(2)}_{i,j})^\alpha [u_i(g)^\dagger \otimes u_j(g)^\dagger] = (h^{(2)}_{i,j})^\alpha,
\end{equation}
[see Fig.~\ref{fig:ph2_A_B}(b)].
This implies that 
\begin{equation}
\label{eqn:AB_transform}
\begin{split}
(u(g) A^\alpha u (g)^\dagger| &= (A^\alpha |V^\alpha(g)^\dagger, \\
|u(g) B^\alpha u(g)^\dagger) & = V^\alpha(g)|B^\alpha),
\end{split}
\end{equation}
where $V^\alpha(g)$ is a unitary representation for the symmetry group $\mathbb{G}$ [Fig.~\ref{fig:ph2_A_B}(c)]. In other words, under the physical transformation, $(A^\alpha|$ and $|B^\alpha)$ transform like the bra and ket vectors in the virtual Hilbert space, respectively. 

{Interestingly,} for several physical models, $(h^{(2)}_{i,j})^\alpha,\ \alpha = 0,1,...,\infty$ are not linearly independent of one another.
In other words, one can find an integer $n$ such that for any $m \geq n$, $ (A^m|B^m)$  is not linearly independent of $\{ (A^\alpha|B^\alpha) \}$ for $\alpha = 0,1,...,n-1$.
This \textit{closed condition} indicates that any polynomial function of $h^{(2)}_{i,j}$, i.e., $P\big(h^{(2)}_{i,j}\big)  = \sum_{m = 0}^{\infty} p_m \big(h^{(2)}_{i,j}\big)^m $, can be written as linear combination of $(A^\alpha|B^\alpha)$ for $\alpha = 0,1,...,n-1$ with coefficients $c_\alpha$:
%
%
%
\begin{equation}
\label{eqn:p_h2_AB}
P\big(h^{(2)}_{i,j}\big)  = \sum_{\alpha=0}^{n-1} c_\alpha (A^\alpha |B^\alpha).
\end{equation}
For example, one can show that all the Hamiltonians composed of the Pauli matrices satisfy Eq.~(\ref{eqn:p_h2_AB}) using the relation $\sigma^i \sigma^j = \delta^{ij}I + i\epsilon^{ijk}\sigma^k $.
The benefit of the closed condition is that Eq.~(\ref{eqn:p_h2_AB}) implies that the local projector in Eq.~(\ref{eqn:expand_h}) can be written as the linear combination of $(A^\alpha|B^\alpha), \alpha = 0,...,n-1$.  
Therefore, if the Hamiltonian is frustration-free, one can always obtain the exact ground state projector by variationally optimizing $n$ parameters $c_\alpha$.
In Sec.\ref{sec:GMPO} and Sec.\ref{sec:GPEPO}, we will show that GTNO can always represent the exact projector to the ground-state subspace of a frustration-free Hamiltonian.
On top of that, GTNO allows accurate description for the non frustration-free systems and can be used to distinguish different quantum phases.

Before explicitly constructing the GTNO using $(A^\alpha|$ and $|B^\alpha)$ in the next section, we provide two concrete examples of the above decompositions.
\\

\noindent
\textbf{Case I: Ising model.}
The local interaction of the Ising model is of the form
\begin{equation}
\label{eqn:ising_h}
h_{i,j}^{(2)} = \sigma^z_i \sigma^z_j,
\end{equation}
which is invariant under the global $\mathbb{Z}_2$ transformation: $\hat U = \prod_i \sigma^x_i$. 
Since $(h_{i,j}^{(2)})^{2m} = I_i I_j$ and $(h_{i,j}^{(2)})^{2m+1} = \sigma^z_i  \sigma^z_j$, we deduce that $n = 2$ with 
\begin{equation}
\label{eqn:ising_AB}
\begin{split}
(A^0| & = (B^0| = \begin{bmatrix} I\end{bmatrix}, \\
(A^1| & = (B^1| = \begin{bmatrix} \sigma^z \end{bmatrix}.
\end{split}
\end{equation}
Given the knowledge of $|A^\alpha)$ and $|B^\alpha)$, we can identify the virtual unitary representation $V^\alpha$ as $V^0 = 1$ and $V^1 = -1$ by direct computations: $|\sigma^x A^0 \sigma^x ) = |A^0)$ and $|\sigma^x A^1 \sigma^x ) = -|A^1)$.
\\

\noindent
\textbf{Case II: Heisenberg model.}
The local interaction of spin-half Heisenberg model is of the form
\begin{equation}
\label{eqn:heisenberg_h2}
h_{i,j}^{(2)} =  \sigma^x_i \sigma^x_{j}  +  \sigma^y_i \sigma^y_{j} + \sigma^z_i \sigma^z_{j},
\end{equation} 
which is invariant under the global $SU(2)$ transformation: $\hat U^\gamma(\theta) = \prod_i  u_i^\gamma(\theta)= \prod_i e^{-i\sigma^\gamma_i\theta/2}$  with $\gamma=[x,y,z]$. 
Since $\sigma^i \sigma^j = \delta^{ij}I + i\epsilon^{ijk}\sigma^k $, we find that $n = 2$ with 
\begin{equation}
\label{eqn:heisenberg_AB}
\begin{split}
(A^0| & = (B^0| = \begin{bmatrix} I\end{bmatrix}, \\
(A^1| & = (B^1| = \begin{bmatrix} \sigma^x & \sigma^y & \sigma^z \end{bmatrix}.
\end{split}
\end{equation}
By direct computation of $|u^\gamma(\theta) A^0 u^\gamma(\theta)^\dagger)$ and $
|u^\gamma(\theta) A^1 u^\gamma(\theta)^\dagger)$, we identify $V^0 = 1$ and $V^1 = e^{-iJ^\gamma_3\theta}$, where $J^\gamma_3$ is the generator of the three-dimensional irreducible $SU(2)$ representation. 

\section{GMPO}
\label{sec:GMPO}
\subsection{{General Construction}}
\begin{figure}
\centering
\includegraphics[width=\linewidth]{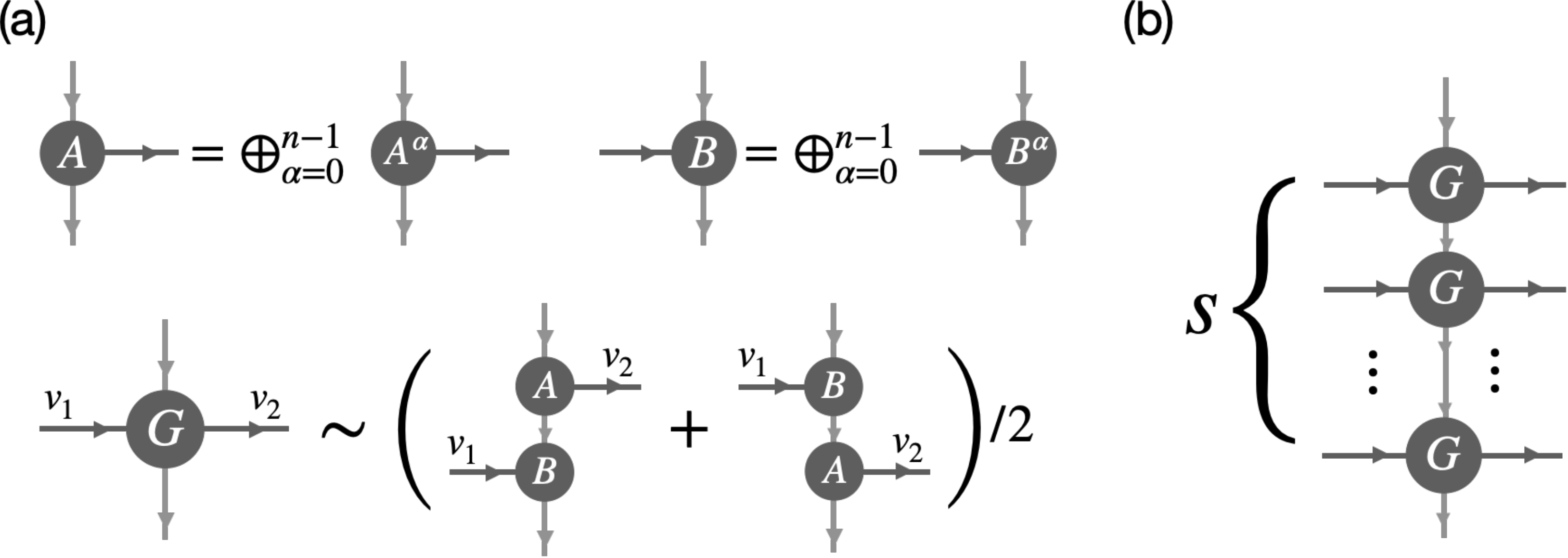}
\caption{(a) The constructions of GMPO given the pairs of $(A^\alpha|$ and $|B^\alpha)$. 
Note that the outgoing and incoming bonds are always attributed to the three-leg tensors $(A|$ and $|B)$, respectively.
(b) The representation power of GMPO can be improved by applying it several times and  {assigning} independent parameters for all the blocks of $A^\alpha B^{\alpha'}$. } 
\label{fig:GMPO_construction}
\end{figure}

We now demonstrate how to construct the 1D GTNO, i.e., the GMPO, given the knowledge of $(A^\alpha|$ and $|B^\alpha)$, $\alpha = 0,...,n-1$.
Let $(A|= \oplus_{\alpha = 0}^{n-1} (A^\alpha|  $ and $|B) = \oplus_{\alpha = 0}^{n-1} |B^\alpha)$, the GMPO with bond dimension $D = \sum_{\alpha = 0}^{n-1} D_\alpha $ can be constructed as 
 \begin{equation}
 \label{eqn:GMPO}
 \begin{split}
  G_{v_1 v_2} & =  \sum_{s_1 s_2}G^{s_1s_2}_{v_1v_2}|s_1\rangle \langle s_2| \\
  & \sim   \sum_{s_1,s_2} \big(A^{s_1 s'}_{v_2} B^{s's_2}_{v_1} +   B^{s_1s'}_{v_1} A^{s' s_2}_{v_2}\big)/2! |s_1\rangle \langle s_2| \\
  & =   \overline{A_{v_1}B_{v_2}}/2!,
 \end{split}
 \end{equation}
 where $\overline{A_{v_1}B_{v_2}}$ denotes the fully symmetrized product. 
{
This can be schematically represented in Fig.~\ref{fig:GMPO_construction}(a).
} 
Here we use the symbol "$\sim$" to emphasize the fact that GMPO is not exactly equal to $\overline{A_{v_1}B_{v_2}}/2!$ but contains $n^2$ variational parameters for each block of $\overline{ A^{\alpha}_{v_1} B^{\alpha'}_{v_2}}/2$ with ${\alpha},{\alpha'} = 0,...,n-1$ in general.
Note that the use of the fully symmetrized product is to take the non-commuting property of different local interactions $h_{i,j}^{(2)}$ into account, which acts trivially if the Hamiltonian is frustration-free.
In other words, if $[h_{i,j}^{(2)}, h_{i',j'}^{(2)}] = 0, \forall i,j,i',j'$, then $\overline{ A^\alpha_{v_1} B^{\alpha'}_{v_2}}/2! = A_{v_1}B_{v_2} = B_{v_2}A_{v_1},\ \forall v_1,v_2$. This implies that with proper choice of the free parameters, GMPO can represent the projector to the ground state subspace of the frustration-free Hamiltonian. 

Besides, since $(A^\alpha|$ and $|B^\alpha)$ transform like the bra and ket vectors in the virtual Hilbert space under the on-site physical transformation [see Eq.~(\ref{eqn:AB_transform})], it is straightforward to show that
 \begin{equation}
 \begin{split}
 \sum_{s_1,s_2} \big(u(g)^{s_1,s'}  G^{s's''} & {(u(g)^\dagger)}^{s'',s_2}\big)  |s_1\rangle  \langle s_2| \\
 = \sum_{s_1,s_2} & V(g)G^{s_1,s_2} V(g)^\dagger |s_1\rangle \langle s_2|,\ \forall g\ \text{in}\ G,
  \end{split}
 \end{equation}
 where $V= \oplus_{\alpha = 0}^{n-1}V^\alpha$. 
 In terms of the TN representation,
 \begin{equation}
 \label{eqn:GMPO_pvrelation}
\includegraphics[width=0.45\linewidth]{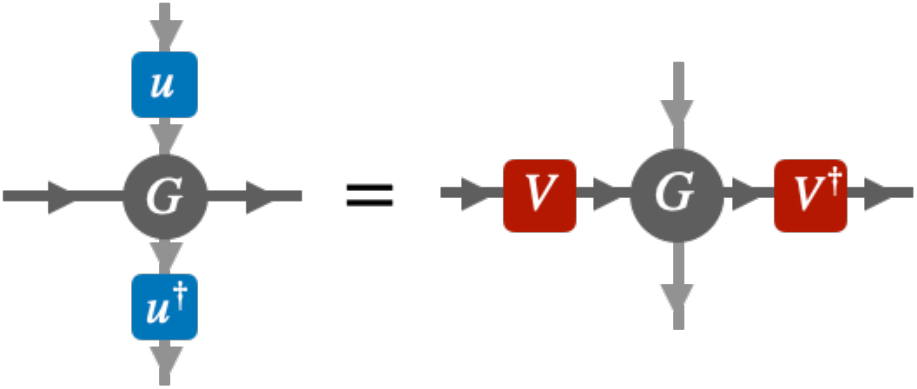}.
\end{equation}
The implication of Eq.~(\ref{eqn:GMPO_pvrelation}) can be understood by considering the global physical operator $\hat G$ formed by contracting all the virtual indices of $G$: 
$\hat G =  \sum_{\bold{s',s}}\cdots G_{}^{s'_{j} s_{j}}  G_{}^{s'_{j+1} s_{j+1}} \cdots | \bold s'\rangle  \langle \bold s |$, where $\bold s = (\cdots ,s_{j}, s_{j+1}, \cdots)$. 
%
%
Since the virtual matrices $V(g)$ cancel out when contracting all the virtual bonds, one can easily see that the global GMPO $\hat{G}$ commutes with the global on-site symmetry operation $[\hat{G}, \hat{U}] = 0$.
In other words, GMPO inherits all the on-site symmetries of the Hamiltonian.

\begin{figure*}
 \centering
\includegraphics[width=1\linewidth]{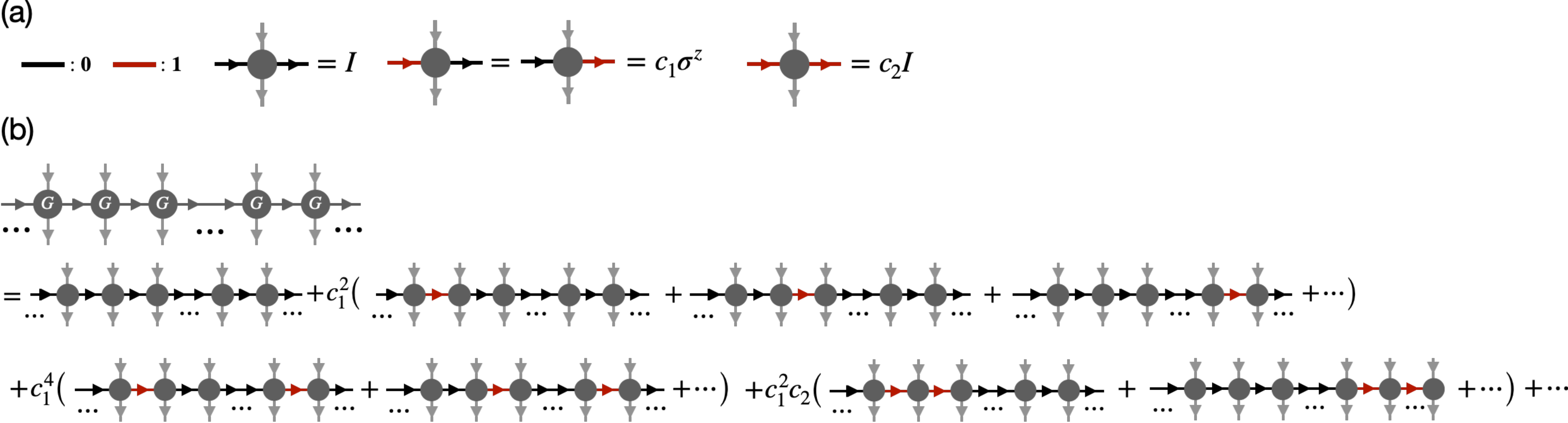}
\caption{(a) Schematic definition of the GMPO for the Ising model. Here we use the black(red) bond to denote the virtual index $0(1)$. (b) The global physical operator $\hat{G}$ generates several string configurations, which represent the physical actions arising from the power of the global Hamiltonian $\hat{H}$. Note that the physical actions of different string configurations follow the rule in (a).}
\label{fig:ising_physical_picture}
\end{figure*}

{Aside from the on-site symmetries, we would like to require GTNO to respect all the symmetries of the Hamiltonian, as the spirit of GTNO is to construct a variational operator generated from the Hamiltonian.
This also allows us to further reduced the number of parameters.}
For example, from  Eq.~(\ref{eqn:ising_AB}), we find that $(A| = (B| =\left[\begin{array}{c|c}
I &  \sigma^z
\end{array} \right]$ for the Ising model. {Here we use the vertical line to distinguish operators arising from different $\alpha =0,...,n-1$.} Enforcing the inversion symmetry and the normalization condition, the GMPO with two variational parameters $c_1,c_2$ can be written as 
\begin{equation}
\label{eqn:ising_GMPO_form}
G= 
\left[
\begin{array}{c|c}
I & c_1\sigma^z \\
\hline
c_1\sigma^z & c_2I
\end{array}\right],
\end{equation}
or equivalently, $G_{00} = I,\ G_{10} =G_{01} = c_1 \sigma^z,\ G_{11} = c_2 I$ [Fig.~\ref{fig:ising_physical_picture}(a)]. 
%
%
To get more intuitive understanding of Eq.~(\ref{eqn:ising_GMPO_form}), we represent the corresponding global operator $\hat G$ schematically in Fig.~\ref{fig:ising_physical_picture}(b) by denoting the virtual index $0(1)$ as the black(red) \textit{string}.
%
One can observe that different types of string correspond to different physical actions generated from the power of $h_{i,j}^{(2)}$. 
Since $[ h_{i,j}^{(2)} ,  h_{i',j'}^{(2)} ] \neq 0$ if $i' = i$ or $j' = j$ in general, GMPO symmetrizes the physical action locally through $ \overline{A_{v_1}B_{v_2}}/2!$ when different strings are connected.
The resulting global operator $\hat{G}$ then generates several physical operators, \textit{both local and non-local}, arising from the power of global Hamiltonian $\hat H^m$ for some $m$. 
The parameters then correspond to the weights of the different connections of strings.

Nonetheless, the current GMPO constructed using Eq.~(\ref{eqn:GMPO}), which we term the {simple} GMPO, cannot generate all the terms existing in the power of $\hat{H}$. 
This is manifested by considering the case of the Heisenberg model.
{From Eq.~(\ref{eqn:heisenberg_AB}), we get $|A) = |B) = \left[ \begin{array}{c|ccc}
I& \sigma^x &  \sigma^y  & \sigma^z \end{array} \right] $. }
The GMPO with bond dimension $D = 4$ and two parameters can then be written as (imposing the inversion symmetry and the normalization condition)
\begin{equation}
\label{eqn:GMPO_Heisenberg}
G = \left[
\begin{array}{c|ccc} 
I  & c_1\sigma^x &  c_1\sigma^y  & c_1\sigma^z \\
  \hline 
|c_1|\sigma^x  & c_2 I & 0 & 0 \\
|c_1|\sigma^y & 0 &  c_2 I& 0\\
|c_1|\sigma^z & 0 & 0 & c_2I\\
\end{array} 
\right], 
\end{equation}
where we allow different signs for the $G$ and $G^T$ to capture both the ferromagnetic and anti-ferromagnetic interaction.
The vanishing of the off-diagonal components in the second diagonal block upon the fully symmetrized product is due to the anti-commutation relation for the Pauli matrices. 
This {reflects} the fact that when considering the power of $\hat H$, $ h_{i,j}^{(2)} h_{j,k}^{(2)}$ always appear  equal number of times as $ h_{j,k}^{(2)} h_{i,j}^{(2)} $, and thus the terms like $\sigma^x_i (\sigma^x_j \sigma^y_j) \sigma^y_k$ will never exist.
However, the simple GMPO cannot generate some long-range physical actions like $\sigma^x_i (\sigma^x_j \sigma^y_j) (\sigma^y_k \sigma^z_k)\sigma^z_l$ which do exist from the power of $\hat{H}$.
To include those terms in our variational ansatz, one can directly apply the simple GMPO several times (say, $s$ times) {as shown in} Fig.~\ref{fig:GMPO_construction}(b). 
The resulting GMPO has bond dimension $D = \big(\sum_{\alpha = 0}^{n-1} D_\alpha  \big)^s$ with $n^{2s}$ variational parameters.
Schematically, this will introduce more string configurations generated from the power of $\hat H$.

Another way to enhance the representation power of GMPO is to increase the number of times each $A^\alpha$ and $B^\alpha$ appears (say, $t_\alpha$ times): 
\begin{equation}
\begin{split}
|A) = &\oplus_\alpha \big[ |A^\alpha)_1\oplus|A^\alpha)_2\oplus \cdots \oplus |A^\alpha)_{t_\alpha}\big]\\
|B) = &\oplus_\alpha \big[ |B^\alpha)_1\oplus |B^\alpha)_2\oplus \cdots \oplus |B^\alpha)_{t_\alpha}\big]
\end{split}.
\end{equation}
The resulting GMPO constructed using Eq.~(\ref{eqn:GMPO}) then have bond dimension $D = \sum_{\alpha = 0}^{n-1} t_\alpha D_\alpha $ with $\big(\sum_{\alpha = 0}^{n-1} t_\alpha\big)^2$ variational parameters.
Physically, this introduces more independent weights for different string configurations. 
For example, by introducing $|A^1) $ and $|B^1)$ two times for the Ising model, the GMPO becomes 
\begin{equation}
G= 
\left[
\begin{array}{c|c|c}
I & c_1\sigma^z & c_3\sigma^z\\
\hline
c_1\sigma^z & c_2I & c_4 I\\
\hline
c_3\sigma^z & c_4I & c_6 I\\
\end{array}
\right].
\end{equation}
Regarding both the virtual index $1$ and $2$ as the red string (since they represent the same physical action), it is then obvious that the additional parameters allow us to assign more independent coefficients for the string configurations in Fig.~\ref{fig:ising_physical_picture}(b).

In the following, we choose to apply the  {simple} GMPO several times and assign independent parameters to each GMPO for simplicity. 
In other words, for the $s$-th order variational GMPS $|\psi^{(s)}\rangle = \hat{G}^s |\psi_0\rangle$ with some initial state $|\psi_0\rangle$, the total GMPO has  bond dimension $D = \big(\sum_{\alpha = 0}^{n-1} D_\alpha  \big)^s$ with $s \times n^2$ free parameters. {We will show that even with such a simple construction, it is already adequate in reflecting the ground-state properties.}
Note that if $\hat{U}|\psi_0 \rangle = |\psi_0\rangle$, the resulting GMPS $|\psi^{(s)} \rangle$ will also respect the on-site symmetry since $[\hat{G}, \hat{U}] = 0$.
This turns out to be a very efficient way to construct the symmetric TN wave functions.

Since the spirit of GTNO {lies in} superpoisng several operators generating from the Hamiltonian, what really matters is the fact that the building-block tensors $(A^\alpha|$ and $|B^\alpha)$ satisfy Eq.~(\ref{eqn:AB_transform}) and Eq.~(\ref{eqn:p_h2_AB}).
{Hence, we generalize the meaning of the superscript $\alpha$ and it no longer needs to represent the power of $h_{i,j}^{(2)}$.}
For example, given the XXZ model $h_{i,j}^{(2)} =  \sigma^x_i \sigma^x_{j}  +  \sigma^y_i \sigma^y_{j} + \Delta \sigma^z_i \sigma^z_{j}$ respecting the $U(1)$ symmetry, the natural choice is $(A^0| = (B^0| = \begin{bmatrix}I\end{bmatrix}$, $(A^1| = (B^1| = \begin{bmatrix} \sigma^x &  \sigma^y  \end{bmatrix} $, and $(A^2| = (B^2| = \begin{bmatrix}\sigma^z\end{bmatrix}$. 
The resulting GMPO is 
\begin{equation}
\label{eqn:GMPO_XXZ}
G = \left[
\begin{array}{c|cc|c} 
I  & c_1\sigma^x &  c_1\sigma^y  & c_3\sigma^z \\
  \hline 
|c_1|\sigma^x  & c_2 I & 0 & 0 \\
|c_1|\sigma^y & 0 &  c_2 I& 0\\
\hline
|c_3|\sigma^z & 0 & 0 & c_4I\\
\end{array} 
\right],
\end{equation}
which has exactly the same form as the Heisenberg model but with more variational parameters.
This is reasonable since XXZ model has less symmetries than the Heisenberg model, and hence its GMPO is less restrictive.

\subsection{Adding One-Site Terms}
\label{subsec:on_site}
Now, we consider the more general Hamiltonian by adding the one-site terms: $\hat H = \sum_{\langle i,j\rangle} h_{i,j}^{(2)} + \sum_i h_{i}^{(1)} $. 
While one can regard $h^{(1)}_i$ as the special case of $h_{i,j}^{(2)}$, here we treat them as different objects, which allows us to assign more variational parameters to the ansatz.
The generalization is straightforward: consider the $\beta$-th power of $h^{(1)}_{i}$, which corresponds to a two-leg tensor, i.e., a physical one-site operator, $C^\beta$. 
If the Hamiltonian is invariant under the on-site symmetry transformation: $\hat U(g) = \prod_{i} u_i(g)$, then
\begin{equation}
u(g) C^\beta u(g)^\dagger =  C^\beta.
\end{equation} 
In other words, under the physical transformation, $C^\beta$ transforms like a scalar in the virtual Hilbert space.
If $h^{(1)}_{i}$ is made out of the Pauli matrices, any polynomial function of $h^{(1)}_{i}$, $Q\big(h^{(1)}_{i}\big)  = \sum_{m = 0}^{\infty} q_m \big(h^{(1)}_{i}\big)^m$, can be expanded using the set of two-leg tensors $C^\beta = \sum_{s_i s'_i}(C^{s_is'_i})^{\beta}|s_i\rangle \langle s'_i|$ ($\beta = 0,1,\cdots,n_o-1$) as 
\begin{equation}
Q\big(h^{(1)}_{i}\big)  = \sum_{\beta=0}^{n_o-1} c_\beta   C^\beta.
\end{equation}
For example, the one-site {transverse} magnetic field $h^{(1)}_i = \sigma^x_i$ corresponds to $n_o = 2$ with $C^0 = I$ and $C^1 = \sigma^x$.
The GMPO can then be constructed as 
\begin{equation}
 \label{eqn:GMPO_onsite}
  G_{v_1 v_2}  \sim   \sum_{\beta = 0}^{n_o-1} \overline{A_{v_1}B_{v_2} C^\beta}/3!.
\end{equation}
In terms of the TN representation, the left hand side of the Eq.~(\ref{eqn:GMPO_onsite}) corresponds to
 \begin{equation}
 \label{eqn:GMPO_onsite_figure}
\includegraphics[width=0.95\linewidth]{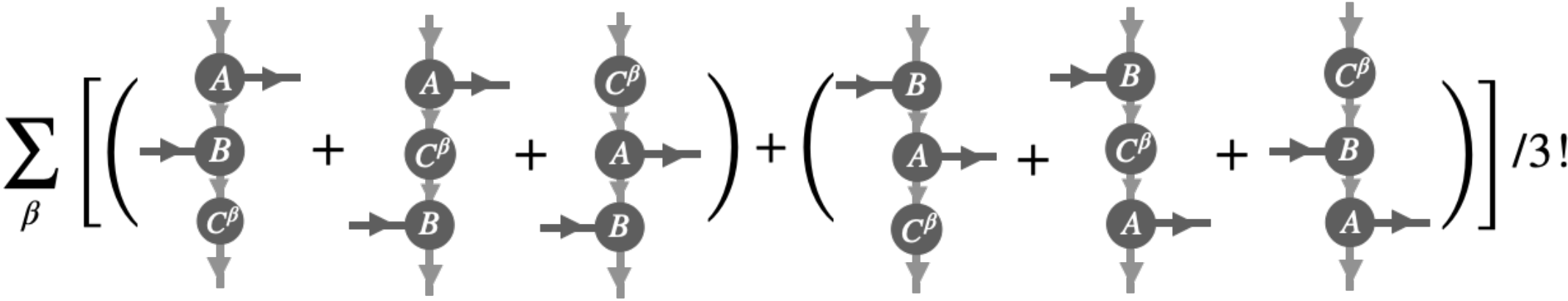}.
\end{equation}

The $ n^2 \times n_o$ variational parameters can then be assigned for each block of $A^\alpha B^{\alpha'} C^\beta$, where $\alpha,{\alpha'} = 0,...,n-1$ and $\beta = 0,...,n_{o}-1$. 
Note that since $C^\beta$ transforms trivially under the physical transformation, the virtual gauge transformation matrix $V(g)$ is fully determined by $|A)$ and $|B)$.
{
Another useful point in the practical simulations is that one can easily write down a program to construct the GMPO using Eq.~(\ref{eqn:GMPO_onsite}) without even knowing the explicit form of it.  
%
}


{
From Sec.\ref{subsec:1D_Heisenberg} to Sec.\ref{subsec:1D_TFCM}, we will use GMPO to study the 1D Heisenberg, transverse field Ising (TFI), and transverse field cluster (TFC) models. 
Specifically, we construct the symmetric GMPS by applying the GMPO to some simple initial states respecting the symmetries of the Hamiltonian, and then variationally optimize the parameters of GMPS to search for the ground states. 
Here we emphasize that GMPS is not merely a class of symmetric TNS \citep{Jiang_2015}. 
The idea of the latter is to sort out all the possible classes of the TNS given the symmetry requirements. 
After the classification, one should variationally optimize all the classes to nail down the phases of the system. 
In contrast, the main objective of the GMPS is not the symmetries but the Hamiltonian of the given system. 
Constructing the GMPO directly from the Hamiltonian and then applying it to some symmetric initial state allows us to obtain the symmetric ansatz effortlessly, bypassing the tedious classification of the symmetric TNS. 
Furthermore, as we will demonstrate below, there exist extra symmetries of the local tensor that only \textit{emerge} for some parameter ranges of the GMPS. 
These emergent symmetries serve as the probes of several quantum phase transitions, including the spontaneously symmetry breaking and the symmetry protected topological phase transitions.
}

\subsection{Case I: Heisenberg Model}
\label{subsec:1D_Heisenberg}

{We now use the GMPO of the form in Eq.~(\ref{eqn:GMPO_Heisenberg}) to study the {$S=1/2$} Heisenberg model.}
%
%
Since the system is $SU(2)$-invariant and critical, the ground state cannot be represented exactly as an injective MPS with a finite bond dimension \citep{Sanz_2009}. 
We choose the initial state as a Majumdar-Gosh state \citep{Majumdar_1969} respecting both the translational and $SU(2)$ symmetries, which can be written as a non-injective MPS $|\psi_0\rangle = \sum_{\bold s} (\cdots M^{s_{n-1}} M^{s_n} M^{s_{n+1}})|\bold s \rangle $ with local tensor $M = \sum_s M^s|s\rangle$ of the form
\begin{equation}
M =\begin{pmatrix}
 0 & 0 &  |0\rangle   \\
 0  &  0 & |1\rangle \\
|1\rangle &-|0\rangle &0\\
\end{pmatrix}.
\end{equation}

Applying the GMPO several times then generates the $SU(2)$ symmetric GMPS $|\psi^{(s)}\rangle = \hat{G}^s|\psi_0\rangle$ with $2s$ free parameters.
To find the optimal parameters for the lowest-energy state, we employ the techniques of the differentiable programming tensor network \citep{Liao_2019,Xie_Liu_2020}, which uses the automatic differentiation to compute the gradient of all the parameters with respect to the energy. This allows us to update all the parameters each time we evaluate the energy.
{Our code implementation for the 1D calculations is publicly available at Ref.~\citep{gtno_code}}

Tab.~\ref{table:GMPO_Heisenberg} shows the energy deviation $\delta E = (E_{\text{calc}} -E_{\text{exact}})/|E_{\text{exact}}|$ comparing with the exact solution \citep{Bethe1931ZurTD} and the correlation length for $s = 1,2,3$.
While the Hamiltonian is not frustration-free, GMPO can get reliable $SU(2)$-symmetric ground state with $\delta E= 1.31 \approx 10^{-5}$ and the extremely large correlation length $\xi \approx 205$ unit cells by {optimizing} merely six free parameters. 
This demonstrates that GMPO accurately captures the long-range correlation  of the system.

\begin{table}
\caption{
{GMPO study of the Heisenberg model.}
{
The second column shows the number of parameters (num of params) used for $|\psi^{(s)}\rangle,\ s = 1,2,3$. The third column displays the energy deviation comparing with the exact solution \citep{Bethe1931ZurTD}, while the fourth column shows the correlation length.
}}
\begin{tabular}{ c c c c }
 \hline \hline
 & \text{ num of params } &$ \delta E$ & $\xi$ \\  
   \hline
$|\psi^{(1)}\rangle$ & 2 & 2.58 $\times10^{-3}$ & 13.06  \\ 
$|\psi^{(2)}\rangle$ & 4 & 1.43 $\times  10^{-4}$ & 58.93\\ 
$|\psi^{(3)}\rangle$& 6 &  1.31 $\times 10^{-5}$ & 204.80\\
 \hline \hline
\end{tabular}
\label{table:GMPO_Heisenberg}
\end{table}

\subsection{Case II: Transverse Field Ising Model}
\label{subsec:1D_TFIM}
The Hamiltonian of the TFI model is written as 
\begin{equation}
\hat H = -\sum_{i } \sigma^z_i \sigma^z_{i+1} +h_x \sum_i  \sigma^x_i,
\end{equation}
which enjoys the global $\mathbb{Z}_2$ symmetry $\hat{U} = \prod_i \sigma^x_i$.
This model is exactly solvable by using the Jordan-Wigner transformation and exhibits the phase transition between the polarized and spontaneously symmetry-broken (SSB) phases at $h_x = 1$.
Using Eq.~(\ref{eqn:GMPO_onsite}), the GMPO with bond dimension $D = 2$ composed of $4$ variational parameters can be written as
\begin{equation}
G(c_s)= 
\begin{pmatrix}
I + c_3\sigma^x& c_1\sigma^z \\
c_1\sigma^z & c_2I+ c_4\sigma^x/3
\end{pmatrix}.
\end{equation}

To gain some intuition, we first consider the extreme limits $h_x = 0,\infty$ such that the Hamiltonian is frustration free.
When $h_x \rightarrow \infty$, the ground state is polarized in the $x$ direction and can be obtained by applying the projector
\begin{equation}
P_{\sigma_x} =  \prod_i (I_i +\sigma^x_i)/2
\end{equation}
to any random state $|\psi_0\rangle$ having non-zero overlap with the ground state.
Up to a normalization constant, this corresponds to the GMPO with parameters $c_3 = 1$ and $c_1 = c_2 = c_4 = 0$. 
On the other hand, when $h_x = 0$, the ground state is two-fold degenerate and its subspace can be obtained by applying the projector 
\begin{equation}
P_{\sigma^z \sigma^z}\prod_{\langle i,j \rangle} (I_i I_j + \sigma^z_i \sigma^z_j)/2.
\end{equation}
This corresponds to the GMPO with parameters $c_1 = c_2 = 1$ and $c_3 = c_4  = 0$.
Therefore, GMPO reproduces the projectors to the ground state subspace in the two extreme limits.

\begin{figure}
 \centering
\includegraphics[width=\linewidth]{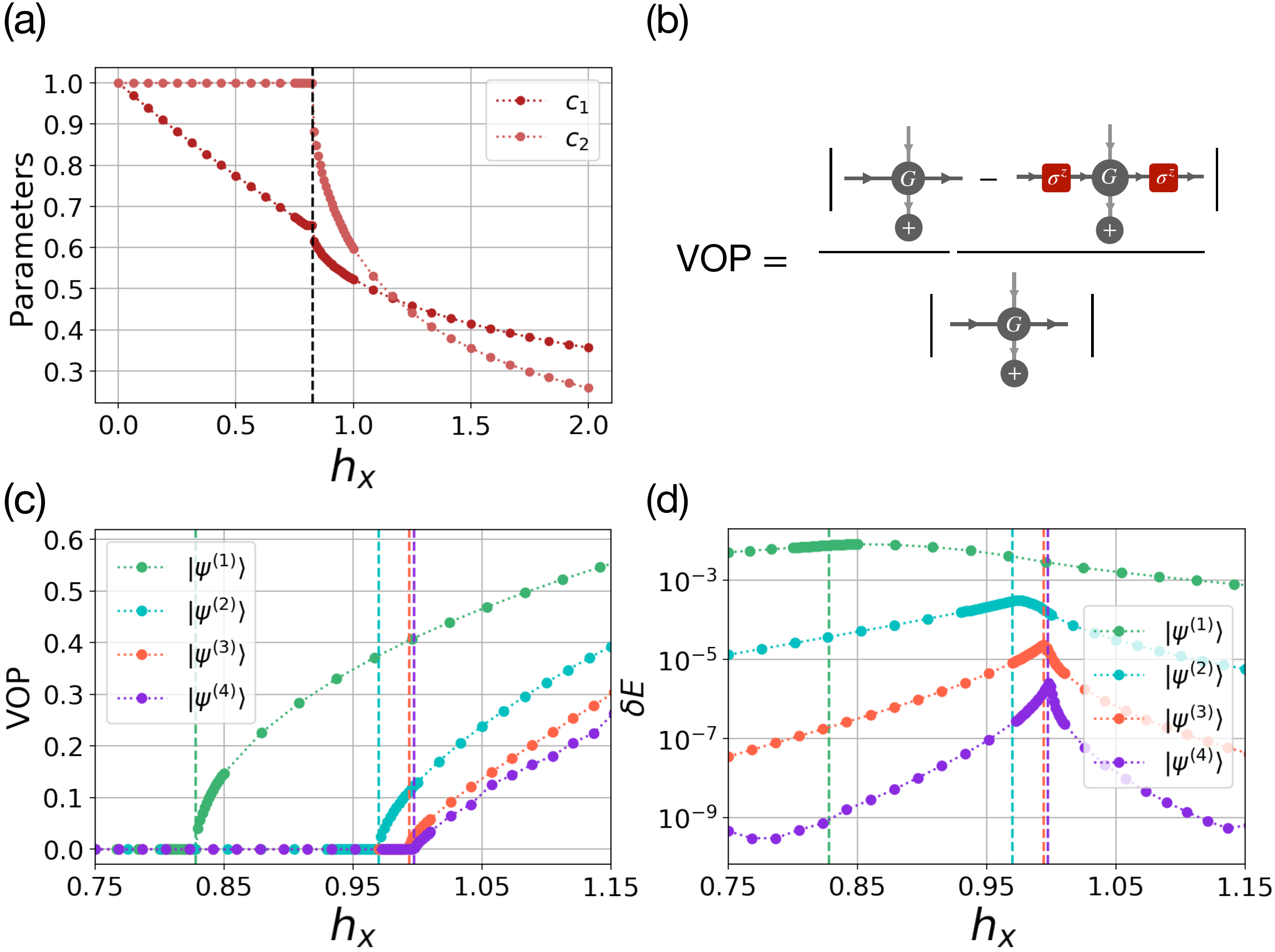}
\caption{GMPO study of the 1D TFI model. (a) The evolution of the variational parameters for $|\psi^{1} \rangle$ as a function of the transverse field $h_x$. (b) The definition of the VOP. (c) The VOP as a function of $h_x$ for $|\psi^{(s)}\rangle,\ s = 1,...,4$. (d) The energy deviation $\delta E$ as a function of $h_x$ for $|\psi^{(s)}\rangle,\ s = 1,...,4$. }
\label{fig:GMPO_TFIM}
\end{figure}

Now, we choose the initial state as the product state polarized in the $x$-direction: $|\psi_0\rangle = \otimes_i |+\rangle_i$, which is the only product state respecting the $\mathbb{Z}_2$ spin-flip symmetry and {having non-zero overlap with the true ground state for all $h_x \geq 0$}. 
The resulting $s$-th order GMPS $|\psi^{(s)}\rangle = \hat{G}^s|\psi_0\rangle$ then also respects the $\mathbb{Z}_2$ symmetry.
For the first-order GMPS $|\psi^{(1)}\rangle$, since $\sigma^x|+\rangle = |+\rangle$, $c_3$ and $c_4$ can be discarded.  
Fig.~\ref{fig:GMPO_TFIM}(a) shows the evolution of $c_1$ and $c_2$ with respect to the transverse field for $|\psi^{(1)}\rangle$. 
One can clearly observe that both of them exhibit a sharp change at $h_x \approx 0.828$, indicating the occurrence of the phase transition. 
To understand the physical meaning of the significant difference between the local tensors above and below $h_x \approx 0.828$, we first consider the limit $h_x = 0$. 
The GMPO yields the Greenberger–Horne–Zeilinger (GHZ) state $|\psi^{(1)}\rangle = \sum_{i = 0}^{1}|i,i,\cdots,i\rangle$ when $c_1 = c_2 = 1$, and the local tensor is of the form
\begin{equation}
M^{(1)} = 
\begin{pmatrix}
|+\rangle & |-\rangle  \\
|-\rangle & |+\rangle 
\end{pmatrix},
\end{equation}
which is non-injective and exhibits a global virtual $\mathbb{Z}_2$ symmetry $W^{(1)} M^{(1)} [W^{(1)}]^\dagger = M^{(1)}$ where $W^{(1)} = \sigma^x$.
In fact, the virtual $\mathbb{Z}_2$ symmetry for MPS is the manifestation of the 1D SSB phase and remains robust for $h_x < h_c$ \citep{Review_TN_2021,Schuch_2010_Ginjective} [see App.~\ref{App:EOP} for details].
Motivated by this extreme limit, we identify the \textit{virtual order parameter} (VOP) as
\begin{equation}
\label{eqn:1d_vop}
v = \frac{| M^{(1)}- W^{(1)} M^{(1)} [W^{(1)}]^\dagger |}{|M^{(1)}|}
\end{equation} 
[see Fig.~\ref{fig:GMPO_TFIM}(b)],
which detects whether the local tensor respects the extra virtual $\mathbb{Z}_2$ symmetry.
{
Here $|\cdot|$ denotes the Frobenius norm of the tensor.
}
Fig.~\ref{fig:GMPO_TFIM}(c) shows the VOP as the function of the transverse field. 
We found that the VOP of $|\psi^{(1)}\rangle$ remains $0$ until $h_x $ increases to around $0.828$, coinciding with the sharp change signaling in the the evolution of the parameters.
Assuming that the virtual $\mathbb{Z}_2$ symmetry is encoded in the first layer of GMPO, the VOP in Eq.~(\ref{eqn:1d_vop}) can be generalized to the higher-order GMPS by choosing $W^{(s)} = I^{\otimes (s-1)} \otimes \sigma^x$.
%
While this may not necessary be the case as the GMPO at different layers commute, we found that it works extremely well if we always {adopt the optimized state $|\psi^{(s)}\rangle$ as the initial parameters for $|\psi^{(s+1)}\rangle$}.
From Fig.~\ref{fig:GMPO_TFIM}(c), one can clearly see that the predicted critical points for the higher-order GMPS are getting closer to the exact point $h_x = 1$.
{
We note that while the VOP remains zero in the SSB phase, its value in the polarized phase {does not} converge to a fixed value when increasing the bond dimensions. 
This shows that the VOP can only be served as a qualitative probe to identify phases and cannot be used to extract the scaling behavior near the phase transition point.
To study the critical behavior, one should adopt a more formal approach by identifying the \textit{entanglement order parameters} {(EOP)} proposed in Ref.~\citep{Iqbal_2021_orderparam}. 
Remarkably, GTNO can also be used to compute the EOP even though we {do not} strongly impose the symmetry constraint of the local tensor as {implemented} in Ref.~\citep{Iqbal_2021_orderparam} (see App.~\ref{App:EOP} for details). 
This demonstrates GTNO's flexibility to combine with other established frameworks.
}
Finally, we compare the variational energy with the exact energy via the relative energy deviation in Fig.~\ref{fig:GMPO_TFIM}(d).
As expected, the relative energy deviation is efficiently lowered by considering the higher-order ansatz. 

\subsection{Case III: Transverse Field Cluster Model}
\label{subsec:1D_TFCM}

The Hamiltonian of the TFC model is written as \citep{Raussendorf_2001, Raussendorf_2003}
\begin{equation}
H = -\sum_j \sigma^z_{j-1} \sigma^x_{j} \sigma^z_{j+1} +h_x \sum_j \sigma^x_j,
\end{equation}
which enjoys the $\mathbb{Z}_2 \times \mathbb{Z}_2=\langle a,b|a^2 = b^2 = e, ab = ba \rangle$  symmetry characterized by the two generators
\begin{equation}
\hat U_{a} = \prod_j \sigma^x_{2j-1},\ \hat U_{b} = \prod_j \sigma^x_{2j},
\end{equation}
with $\hat U_{i} \hat H  \hat U_{i}^\dagger = \hat H, i = a,b$.
This model is exactly solvable and exhibits a phase transition between the symmetry-protected topological and polarized phases at $h_x = 1$ \citep{Doherty_2009}.
Importantly, both phases respect the $\mathbb{Z}_2 \times \mathbb{Z}_2$ symmetry and cannot be distinguished by any local order parameters.
Instead, they should be distinguished by the string order parameters \citep{Kennedy_Tasaki_1992} and the degeneracy of the entanglement spectra \citep{Li_Haldane_2008,Pollmann_Turner_2010}. 
Phrasing in the TN language, the corresponding virtual gauge transformation matrix $V(g)$ of the symmetric MPS should form a non-trivial projective representations of the symmetry group $\hat{G}$ \citep{Pollmann_Turner_2010, Chen_2011_1Dclassification, Schuch_PG_2011}.

While the interaction is three-site, one can treat them as the two-site interactions by blocking two spins into a single unit cell as follows:
\begin{equation}
\label{eqn:cluster}
\begin{split}
&-\sum_j \sigma^z_{j-1} \sigma^x_{j} \sigma^z_{j+1} \\
&= -\frac{1}{2}\sum_j \big( \sigma^z_{j-1} \sigma^x_{j} \sigma^z_{j+1} I_{j+2} + I_{j-2}\sigma^z_{j-1} \sigma^x_{j} \sigma^z_{j+1} \big) \\
&  = -\frac{1}{2}\sum_k (\sigma^z \otimes \sigma^x)_k(\sigma^z \otimes I)_{k+1} + (I \otimes \sigma^z )_k(\sigma^x\otimes \sigma^z )_{k+1} \\
&= \sum h_{k,k+1}. 
\end{split}
\end{equation}
The generators of $\mathbb{Z}_2 \times \mathbb{Z}_2$ symmetry then becomes 
$ \hat U_{a} = \prod_k (\sigma^x \otimes I)_k$, and $\hat U_{b} = \prod (I \otimes \sigma^x)_k$.
From Eq.~(\ref{eqn:cluster}), the sets of $|A^\alpha)$ and $|B^\alpha)$ with $n = 2$ can be easily identified as 
\begin{equation}
\begin{split}
(A^0| & = (B^0| = \begin{bmatrix}I \otimes I\end{bmatrix}, \\
(A^1| & = \begin{bmatrix}
\sigma^z \otimes \sigma^x & I\otimes \sigma^z
\end{bmatrix},\  (B^1| = \begin{bmatrix}
\sigma^z \otimes I & \sigma^x \otimes \sigma^z
\end{bmatrix}, \\
(A^2| & = \begin{bmatrix}
\sigma^z \otimes -i\sigma^y 
\end{bmatrix},\  (B^2| = \begin{bmatrix}
-i\sigma^y \otimes \sigma^z
\end{bmatrix}, 
\end{split}
\end{equation}
with the corresponding virtual transformation matrix
\begin{equation}
\label{eqn:TFCM_Vg}
\begin{split}
[V_{a}]^0 &= [V_{b}]^0 = 1,\\
[V_{a}]^1 & = \sigma^z,\ [V_{b}]^1 = -\sigma^z,\\
[V_{a}]^2 & = 1,\ [V_{b}]^2 = -1.
\end{split}
\end{equation}
Similarly, $C^\beta$ with $n_o = 4$ can be identified as $C^0 = I \otimes I,\ C^1 = I \otimes \sigma^x,\ C^2 = \sigma^x \otimes I,$ and $C^3 = \sigma^x \otimes \sigma^x$.
For simplicity, we let $C^\beta$ with $\beta = 1,2,3$ on each $A^\alpha B^{\alpha'}$ share the same parameter. 
The resulting GMPO after imposing the normalization condition and inversion symmetry will then have $11$ free parameters. 

\begin{figure}
 \centering
\includegraphics[width=\linewidth]{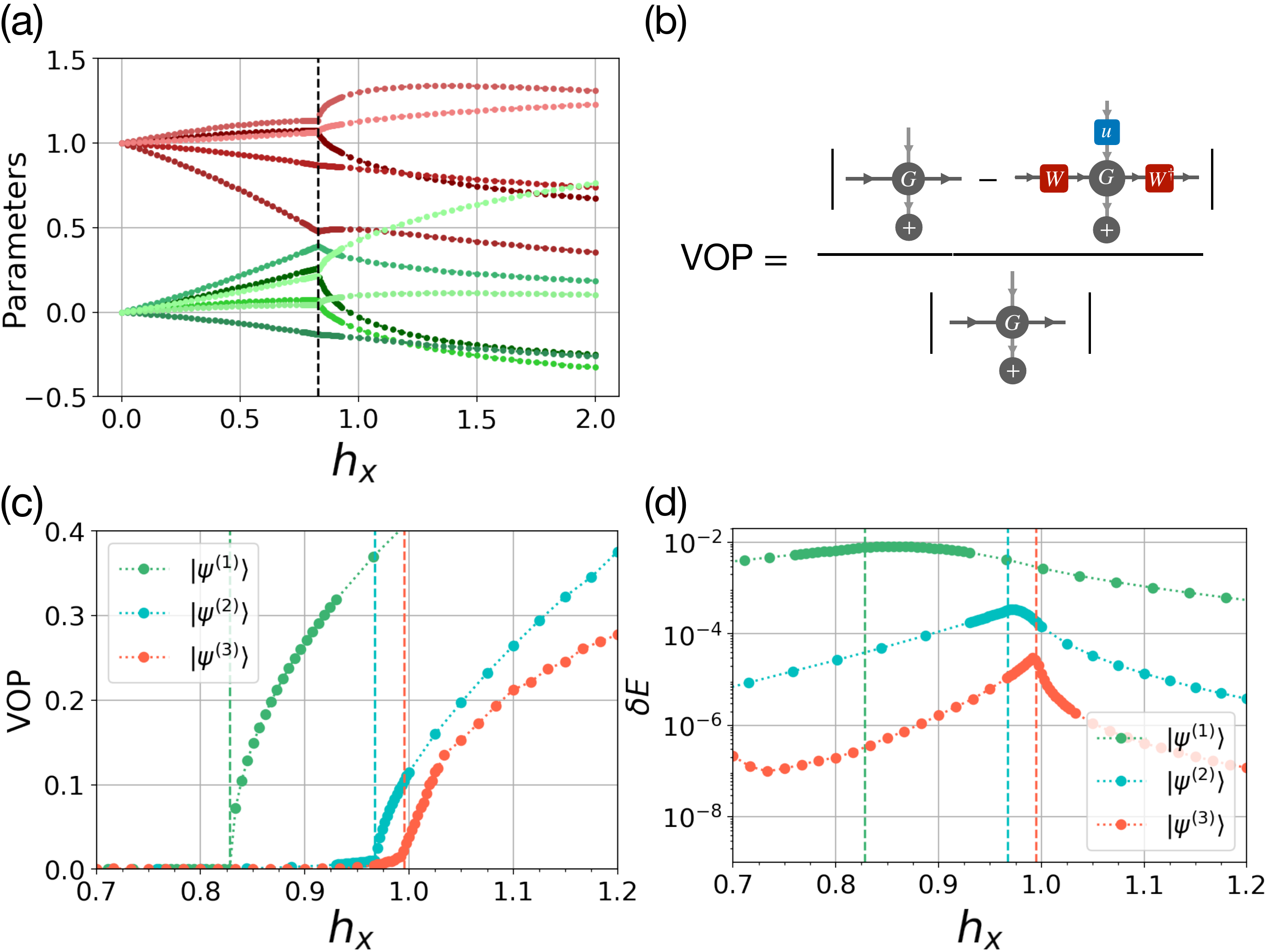}
\caption{GMPO study of 1D TFC model. (a) The evolution of the variational parameters for $|\psi^{1} \rangle$ as a function of the transverse field $h_x$. The red(green) colors denote the parameters attributed to the two-site (one-site) interactions.
(b) The definition of the VOP. (c) The VOP as a function of $h_x$ for $|\psi^{(s)}\rangle,\ s = 1,2,3$. (d) The energy deviation $\delta E$ as a function of $h_x$ for $|\psi^{(s)}\rangle,\ s = 1,2,3$. 
}
\label{fig:GMPO_TFCM}
\end{figure}

Now, we choose the initial state as the product state polarized in the $x$-direction: $|\psi_0\rangle = \otimes_i |+\rangle_i$, which is the only product state respecting the $\mathbb{Z}_2 \times \mathbb{Z}_2 $ spin-flip symmetry and {having non-zero overlap with the true ground state for all $h_x \geq 0$.}
The resulting $s$-th order GMPS $|\psi^{(s)}\rangle = \hat{G}^s|\psi_0\rangle$ then also respects the $\mathbb{Z}_2 \times \mathbb{Z}_2$ symmetry.
Fig.~\ref{fig:GMPO_TFCM}(a) shows the evolution of the variational parameters with respect to the transverse field for $|\psi^{(1)}\rangle$. 
Similar to the TFIM, one can observe that all the parameters exhibit a sharp change at $h_x \approx 0.828$, signaling the occurrence of the phase transition.
To identify the VOP, we first note that from Eq.~(\ref{eqn:TFCM_Vg}), the virtual gauge transformation of GMPO $V^{(i)} = \oplus_\alpha [V^{(i)}]^\alpha$ is the linear representation of the $\mathbb{Z}_2 \times \mathbb{Z}_2$ symmetry. 
Therefore, the corresponding symmetric GMPS describes the topological trivial phases in general.
To see how the non-trivial projective representation emerges in the SPT phase, we consider the zero field limit $h_x = 0$. 
The projector to the ground state subspace can be represented by GMPO with all the parameters for the two-site interaction, i.e., $A^\alpha B^{\alpha'} C^0$, becoming equal, while all the parameters for the one-site interactions, i.e., $A^\alpha B^{\alpha'} C^{\beta \neq 0}$, vanish. 
In this extreme limit, one can show that the GMPO satisfies the following physical-virtual transformation:
\begin{equation}
\includegraphics[width=0.45\linewidth]{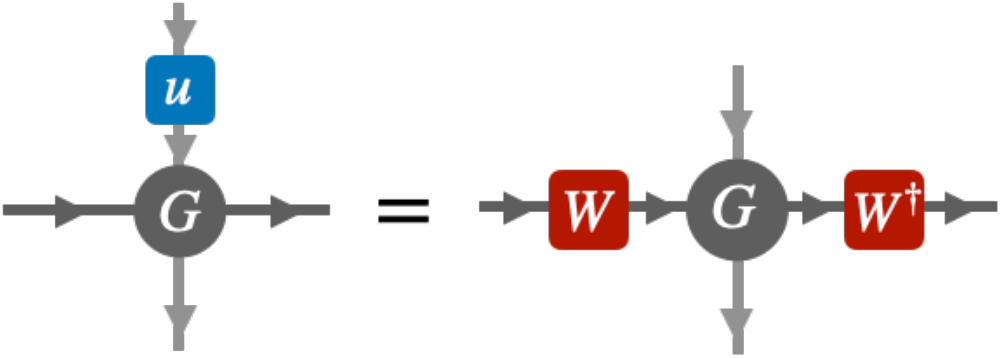},
\end{equation} 
with $W^{a} = \sigma^x\otimes \sigma^z$ and $W^{b} = I\otimes \sigma^x$.
Since $W^{a} W^{b} = -W^{b} W^{a} $, it forms the projective representation of the $\mathbb{Z}_2 \times \mathbb{Z}_2$ symmetry.
The VOP can then be defined as the difference between the original and transformed tensors as shown in  [Fig.~\ref{fig:GMPO_TFCM}(b)].
Using Eq.~(\ref{eqn:GMPO_pvrelation}), one can derive that $W^{a,(s)} = I^{\otimes (2s-2)} \otimes W^a $  for the $s$-th ordered GMPS.
Fig.~\ref{fig:GMPO_TFCM}(c) shows the VOP as the function of $h_x$ for $|\psi^{(s)}\rangle,\ s=1,2,3$. 
The non-zero value of VOP signals the occurrence of the phase transition and the estimated transition point is moving closer to the exact critical point $h_x = 1$ when considering the higher-order GMPS.
We also compute the relative energy deviation, showing that higher order GMPS yields more accurate energy of the ground state [Fig.~\ref{fig:GMPO_TFCM}(d)].

\section{GPEPO}
\label{sec:GPEPO}

In this section, we generalize the GMPO construction to higher dimensions. While we focus on the 2D square lattice,  generalization to other geometries is straightforward.
Given $(A|= \oplus_{\alpha = 0}^{n-1} (A^\alpha|  $ and $|B) = \oplus_{\alpha = 0}^{n-1} |B^\alpha)$, the GPEPO with bond dimension $D = \sum_{\alpha = 0}^{n-1} D_\alpha $ and $n^4$ free parameters can be constructed as 
\begin{equation}
\label{eqn:GPEPO}
G_{v_1 v_2 v_3 v_4} = \overline{ A_{v_1} A_{v_2} B_{v_3} B_{v_4}}/4!
\end{equation}
or schematically, 
\begin{equation}
\includegraphics[width=\linewidth]{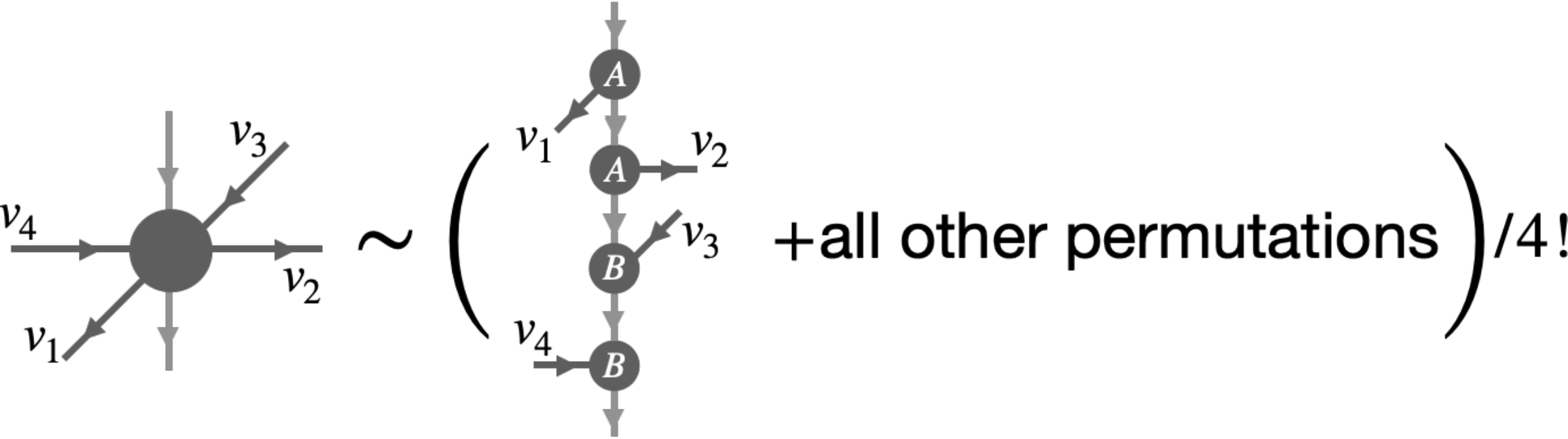}.
\end{equation}
{Similar to the GMPO, the outgoing and incoming bonds are always attributed to $(A|$ and $|B)$, respectively.}
This immediately guarantees that the physical transformation can be pulled through the virtual bonds (using  Eq.~(\ref{eqn:AB_transform})):
\begin{equation}
\label{eqn:GPEPO_pvrelation}
\includegraphics[width=0.45\linewidth]{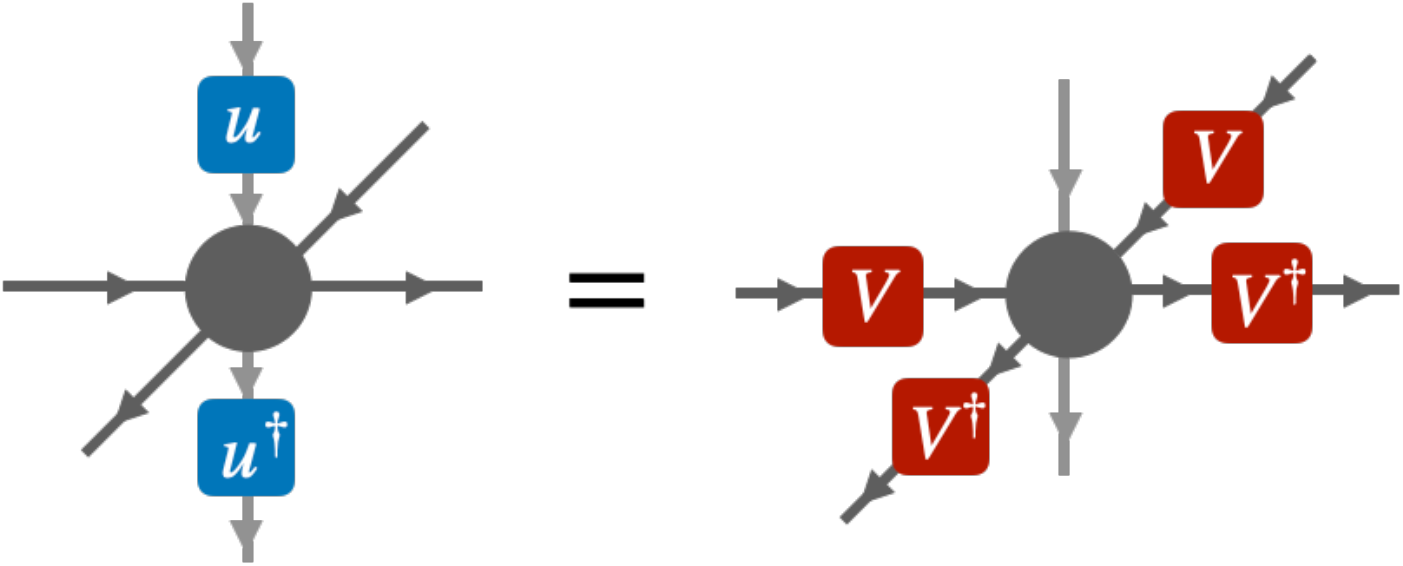}.
\end{equation}
Therefore, the global operator $\hat{G}$ formed by contracting all the virtual bonds of $G_{v_1 v_2 v_3 v_4}$ inherits all the on-site symmetries from the Hamiltonian.
By applying the GPEPO to some simple symmetric PEPS tensor, one can then construct the symmetric GPEPS ansatz to variationally study 2D systems.
The incorporation of the onsite terms is straightforward as well: given the set of two-leg tensors $C^\beta = \sum_{s_i s'_i}(C^{s_is'_i})^{\beta}|s_i\rangle \langle s'_i|\}$ ($\beta = 0,1,\cdots,n_o-1$), we consider the GPEPO
\begin{equation}
\label{eqn:GPEPO_onsite}
G_{v_1 v_2 v_3 v_4} = \sum_{\beta = 0}^{n_o -1}\overline{ A_{v_3} A_{v_4}B_{v_1} B_{v_2} C^\beta }/5!.
\end{equation}
We note that while Eq.~(\ref{eqn:GPEPO_onsite}) seems to be complicate, one can easily write down a program to construct the GPEPO without knowing its explicit form.

{
In Sec.\ref{subsec:2D_Heisenberg}-\ref{subsec:TCXM}, we will use the GPEPO to study the 2D Heisenberg, TFI, and toric code \citep{kitaev_2003} in a magnetic field (TCX) models.
Note that one cannot apply Eq.~(\ref{eqn:GPEPO_onsite}) to the TCX model, as it consists of the four-site interactions.
While we currently have no simple recipe to write down the general four-site GPEPO, we provide explicit construction of the GPEPO for the TXC model in Sec.\ref{subsec:TCXM}.
Besides, we remark that the most unbiased initial state for GTNO is the simplest state respecting all the symmetries of the Hamiltonian. 
As demonstrated in Sec.\ref{sec:GMPO}, the symmetric GMPS can still describe the SSB phases, where the degeneracy of the ground states is encoded in the virtual symmetries.
However, to compare with other numerical approaches for the 2D cases, we will use the prior knowledge about the system to choose the initial state for 2D Heisenberg and TFI models, which is also assumed in Ref.~\citep{10.21468/SciPostPhys.10.1.012} and \citep{Vanderstraeten_2017_perturbation}
}

Different from MPS, PEPS has no exact canonical form and cannot even be contracted exactly.
This renders an efficient optimization of the parameters in a PEPS to be much more challenging than the MPS.
Recent progress has been made to effectively evaluate the gradient of the energy functional either diagrammatically \citep{Corboz_2016_Var, Vanderstraeten_2016_GD} or using the automatic differentiation approach developed in the machine learning community \citep{Liao_2019_Diff}. {We will adopt the latter approach using peps-torch \citep{Hasik} to optimize the ansatz once the GPEPS has been established for the desired models.}
Peps-torch has been demonstrated with a very good capability for various spin systems such as frustrated Heisenberg antiferromagnet \citep{10.21468/SciPostPhys.10.1.012, arXiv:2110.11138}, chiral spin liquid \citep{Hasik_2022}, and many other quantum magnetism \citep{Commun.Phys.5.130, arxiv:2204.01197}, as well as bosonic systems \citep{PhysRevA.102.053306}. 
%

\subsection{Case I: Heisenberg Model}
\label{subsec:2D_Heisenberg}
Now we study the 2D anti-ferromagnetic Heisenberg model using the GPEPO evaluated through Eq.~(\ref{eqn:GPEPO}) with $|A) = |B) = \left[ \begin{array}{c|ccc}
I& \sigma^x &  \sigma^y  & \sigma^z \end{array} \right] $. 
Since the ground state develops Neel order, we choose the initial state as the product state polarized in the $z$-direction $|\psi^{(0)} \rangle = \otimes_i |\uparrow \rangle_i$, {and then rotate the physical $\sigma^z$ basis by unitaries $-i \sigma^y$ at one of the sublattices.}
The resulting GPEPS $|\psi^{(n)} \rangle = \hat{G}|\psi^{(0)}\rangle$ then acquires the $U(1)$ symmetry, the subgroup of $SU(2)$.
To yield more accurate ground states of GPEPO for the given bond dimension $D = 4$, we note that since the ground states is not $SU(2)$ symmetric, one can relieve the $SU(2)$-symmetric constraint on the GPEPO.
Therefore, we consider the $n = 2$ with $(A^0| = (B^0| = \begin{bmatrix}I\end{bmatrix}$, $(A^1| = (B^1| = \begin{bmatrix} \sigma^x &  \sigma^y  \end{bmatrix} $, and $(A^2| = (B^2| = \begin{bmatrix}\sigma^z\end{bmatrix}$. 
The resulting GPEPO constructed using Eq.~(\ref{eqn:GPEPO}) is then only invariant under the $U(1)$ on-site transformation $\hat U(\theta) = \prod_i e^{-i\theta \sigma^z_i/2}$. {Such GPEPO is also feasible for studying the XXZ model.}
%

Tab.~\ref{table:GPEPO_Heisenberg} shows energies and magnetizations obtained by two different GPEPOs and VarOpt. 
Here, the PEPS of VarOpt is also optimized using peps-torch {with the only condition of imposing the $C_{4v}$ symmetry}.
The GPEPSs, with extremely few parameters ($5$ and $16$ for $SU(2)$ and $U(1)$ GPEPO, respectively), yield very close energies and magnetizations compared to the results by VarOpt, which contains $110$ parameters. 
These outcomes provide a clear evidence that {GPEPS faithfully capture the physics of the Heisenberg model using much less parameters.} 

\begin{table}
\caption{
Comparison of energy density $E$ and magnetization $m$ between GPEPO and VarOpt (imposing $C_{4v}$ symmetry) with bond dimension $D = 4$ for the 2D Heisenberg model. The extrapolated Quantum Monte Carlo result is $E = -0.669437(5)$ and $m = 0.3070(3)$ \citep{Sandvik_1997}. 
}
\begin{tabular}{ c c c c}
 \hline  \hline
& num of params & E & m  \\ 
 \hline 
{GPEPS}[$SU(2)$] & 5 &-0.668839 & 0.33972\\ 
{GPEPS}[$U(1)$] & 16 & -0.668949 & 0.33638   \\
VarOpt & 110 & -0.668951 &  0.33640 \\ 
 \hline \hline
\end{tabular}
\label{table:GPEPO_Heisenberg}
\end{table}

\subsection{Case II: Transverse Field Ising Model}
\label{subsec:2D_TFIM}

We now turn to studying the 2D TFI model, whose Hamiltonian reads
\begin{equation}
\hat H = -\sum_{\langle i,j \rangle } \sigma^z_i \sigma^z_{j} +h_x \sum_i  \sigma^x_i.
\end{equation}
Same as in one dimension, this model possesses a global $\mathbb{Z}_2$ symmetry and serves as a good benchmark model for probing the SSB phase. 
However, unlike in 1D, it cannot be exactly solved in two dimension. 
Its critical point can, nevertheless, be well estimated through numerical calculation and earlier Monte Carlo simulation has predicted a critical point at $h_c=3.04438(2)$ \citep{Blote_2002_MCTFIM}.

\begin{figure}
\includegraphics[width=\linewidth]{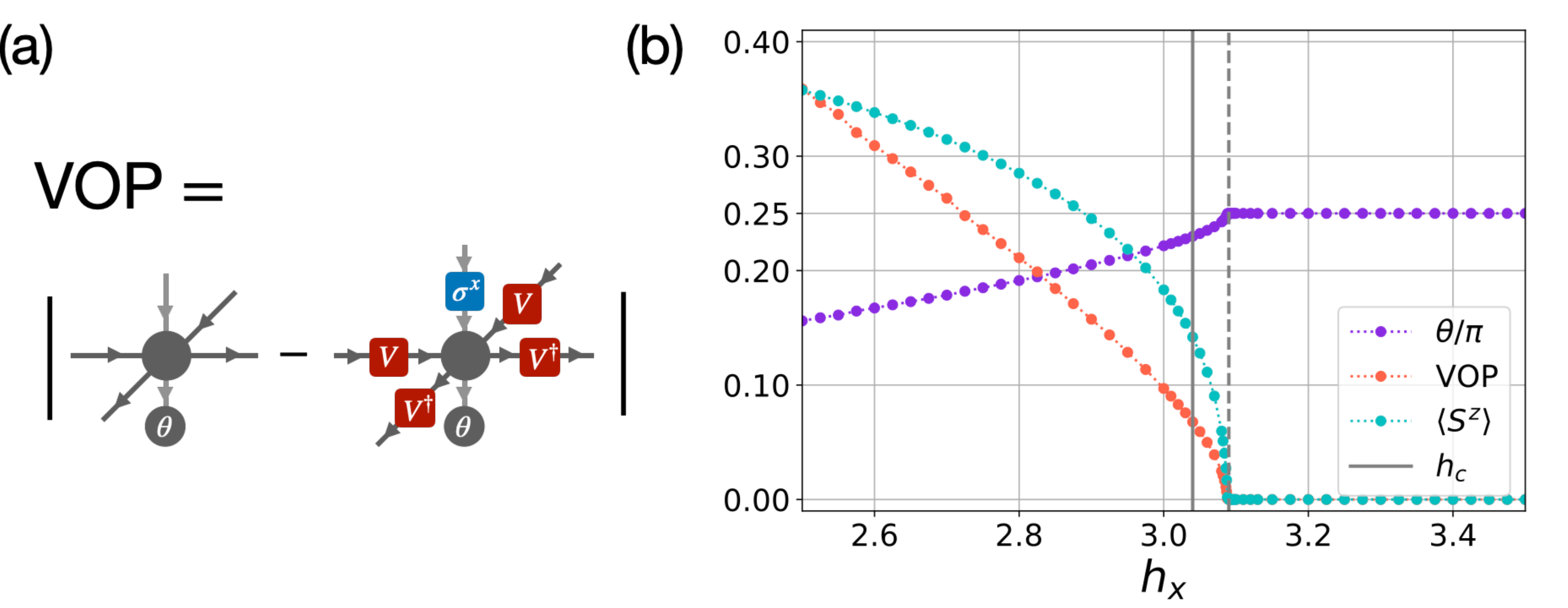}
\caption{GPEPO study of the TFI model.
(a) The definition of VOP. (b) {The parameter} $\theta/\pi$ {of the initial product state}, VOP, and $\langle S^z \rangle$ as a function of $h_x$ for $|\psi^{(1)}\rangle$ {(see the main text for details)}.
The dashed line indicates the estimated critical point and the solid line points out the $h_c$ from Quantum Monte Carlo \citep{Blote_2002_MCTFIM}.
}
\label{fig:GPEPO_TFIM}
\end{figure}

The GPEPO for 2D TFI model can be constructed following Eq.~(\ref{eqn:GPEPO_onsite}) with $|A) = |B) = \left[ \begin{array}{c|c}
I& \sigma^z \end{array} \right] $ and $C^0=I$, $C^1=\sigma^x$. 
For forming the GPEPS, we choose $|\psi_0(\theta)\rangle = \otimes_i ( \cos\theta |\uparrow\rangle_i + \sin\theta|\downarrow \rangle_i$) with $0\leq \theta \leq 0.25\pi$.
Note that in the symmetry-breaking phase, the current GPEPS favors only one of the degenerate ground states.
%
This can be observed by considering the physical action $\sigma^x$ on the local tensor using Eq.~(\ref{eqn:GPEPO_pvrelation}):
\begin{equation}
\label{eqn:TFIM_pvrelation}
\includegraphics[width=0.45\linewidth]{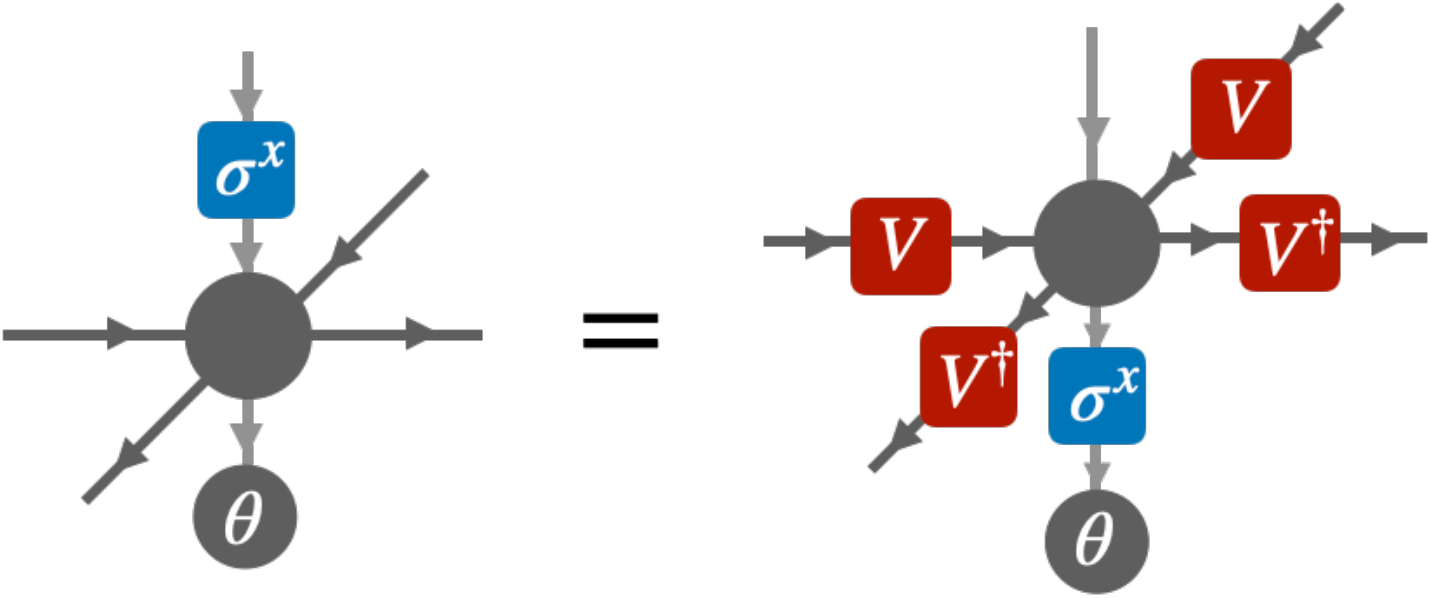},
\end{equation}
where $V = [1] \oplus [-1] = \sigma^z$.
This implies that the GPEPS will respect the global $\mathbb{Z}_2$ symmetry only when the initial state $|\psi_0\rangle$ respects the symmetry, i.e., $\theta/\pi = 0.25$.
Eq.~(\ref{eqn:TFIM_pvrelation}) motivates us to define the VOP as the norm difference between the original and symmetry-transformed tensor [Fig.~\ref{fig:GPEPO_TFIM}(a)]. 
Fig.~\ref{fig:GPEPO_TFIM}(b) shows the evolution of $\theta/\pi$, VOP, and $\langle S^z \rangle$ with the change of $h_x$ for $|\psi^{(1)}\rangle = \hat{G} |\psi_0 \rangle$.
The physical order parameter $\langle S^z \rangle = \langle \sigma^z\rangle /2 $ estimates the phase transition point at $h_x \approx 3.09$.
Besides, the parameter of the initial state $\theta/\pi$ sticks to a fixed value $0.25$ for $h_x \gtrsim 0.309$, consistent with our previous claim that the GPEPS will respect the spin-flip symmetry only when $|\psi_0\rangle$ is invariant under the symmetry. 
The phase transition is also captured by the vanishing of VOP in the symmetry-preserved phase.
We also consider the 2nd order GPEPS, and the results are similar to $|\psi^{(1)}\rangle$ with more accurate estimated critical point at $h_x \approx 0.3051$.  

Tab.~\ref{table:GPEPO_TFI} shows the comparison between GPEPS with previous results using VarOpt \citep{Vanderstraeten_2016_GD} and the perturbation method \citep{Vanderstraeten_2017_perturbation}. 
For $D=2$, GPEPS estimates the same critical point as the VarOpt did, which is far better than the results obtained using the perturbation method.
However, we note that the number of parameters for GPEPS is only slightly lesser than {that of VarOpt}. 
To demonstrate the power of GPEPO, we also consider the 2nd order GPEPS.
With merely $19$ free parameters, GPEPS predicts a more accurate critical point than the VarOpt did for $D = 3$ using $42$ parameters.
%

%

\begin{table}
\caption{
The predicted critical point of the 2D TFI model compared to the VarOpt \citep{Vanderstraeten_2016_GD} and the perturbation method \citep{Vanderstraeten_2017_perturbation}. Numbers in the parentheses represent the number of variational parameters.
The critical point is located at $h_c = 3.04438(2)$ as estimated by the Monte Carlo simulations \citep{Blote_2002_MCTFIM}.
}
\begin{tabular}{c c c c}
 \hline  \hline
& D = 2 & D = 3 & D = 4  \\ 
 \hline 
GPEPS & 3.09 (10) & N/A & 3.051 (19) \\ 
VarOpt & 3.09 (12*) & 3.054 (42*) & N/A \\ 
Perturbation & 3.35 (3) &  3.1 (5) & N/A \\
 \hline \hline
\end{tabular}

\raggedright{*under the assumption that only the $C_{4v}$ symmetry is taken into account.}
\label{table:GPEPO_TFI}
\end{table}

\subsection{Case III: Toric Code in a Magnetic Field}
\label{subsec:TCXM}
\begin{figure}
\includegraphics[width=\linewidth]{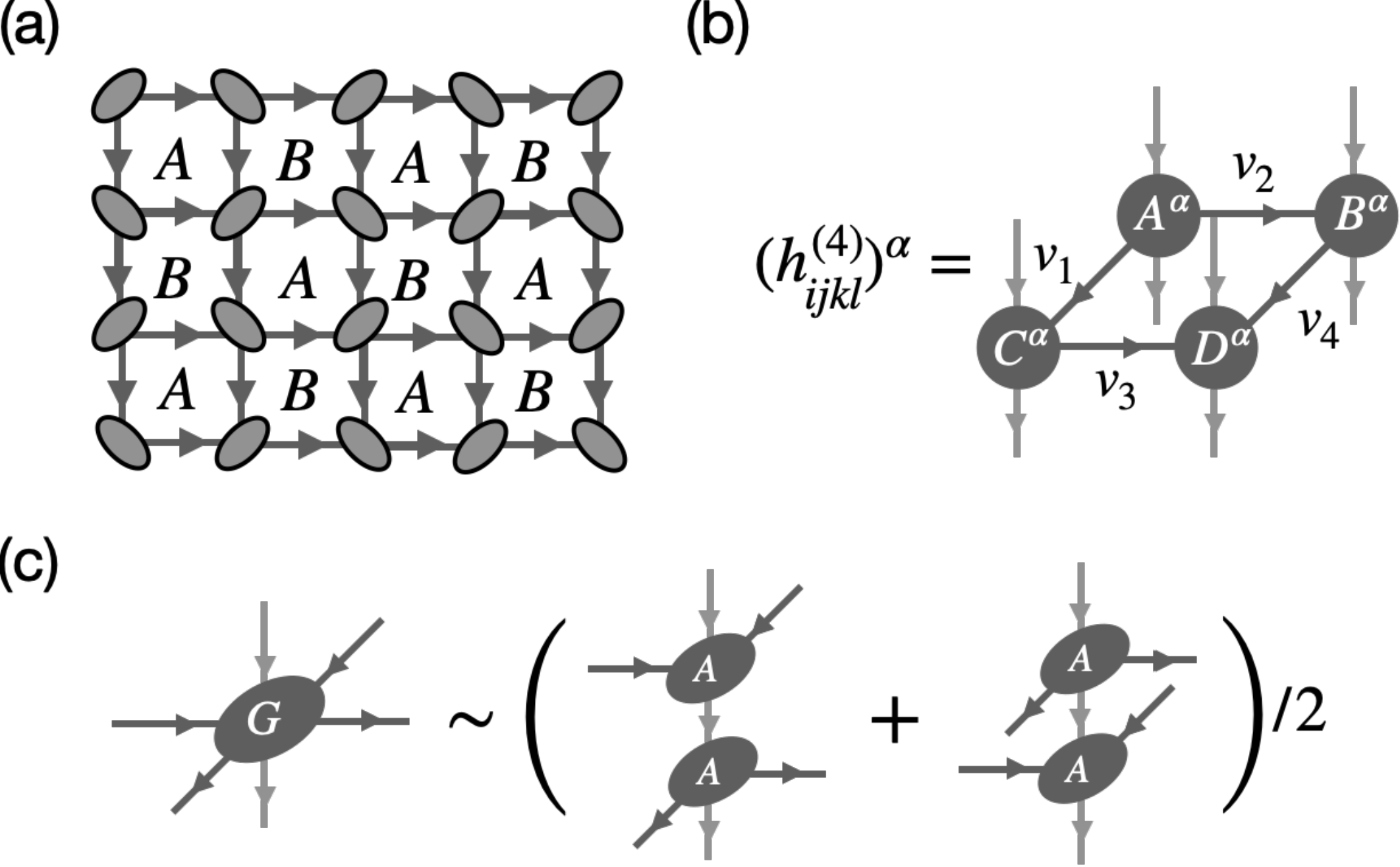}
\caption{(a) The toric code model can be defined on the square lattice associated with two types of plaquette and the spin-$1/2$ degrees of freedom placed on the vertices. (b) The $\alpha$-th power of four-site interaction can be decomposed into the contraction of four four-leg tensors. (c) The construction of GPEPO for the toric code model.} 
\label{fig:toric_code}
\end{figure}

%
The toric code model \citep{kitaev_2003} can be defined on the square lattice associated with two types of plaquette, A and B, and the spin half degrees of freedom placed on the vertices [Fig.~\ref{fig:toric_code}(a)].
The Hamiltonian consists of the four-site interactions acting on the two plaquettes, and can be written as
\begin{equation}
\label{eq:GPEPO_toriccode}
H_\text{TC} = -\sum_{A} h_A  -\sum_{B}  h_{B},
\end{equation}
where $h_A =  \prod_{i \in A} \sigma^z_i$ and $h_B = \prod_{i \in B} \sigma^x_i$. 
Since the model is frustration-free, i.e., $[h_A,h_B] = 0,\ \forall A,B$, the ground state subspace is spanned by the states satisfying $h_A|\psi\rangle = h_B|\psi\rangle =+ |\psi\rangle$.
The system is in the intrinsic topologically ordered phase which is characterized by the topological degeneracies and anyonic quasiparticle statistics \citep{Wen04quantumfield}.
Similar to the SPT phases, topologically ordered phases cannot be identified using any local order parameters. Instead, they should be detected by the topological entanglement entropy, features of the entanglement spectrum, and the modular matrices extracted from the full set of the ground states respecting anyonic excitations \citep{Kitaev_2006_EE,Levin_2006_EE,Cirac_2011_ES,Moradi_2015}.
Recently, it has been shown that the topological order can be classified using the virtual symmetries of the PEPS tensor \citep{Schuch_2010_Ginjective,Williamson_2016_SPT}.

To study the topological phase transition, we subject the model to a magnetic field pointing in the $x$ direction, which we call the TCX model.
The Hamiltonian can then be written as
\begin{equation}
H = H_\text{TC} + h_x \sum_i \sigma^x_i.
\end{equation}
The ground state belongs to the charge-free sector  $\langle h_B \rangle = +1$ for all $h_x \geq 0$.
Using this property, the model can be mapped to the 2D TFI model \citep{Trebst_2007,Hamma_2008}, and thus the phase transtion point can be identified at $1/2.044382 \approx 0.3285$.
On the optimization side, we can focus on the charge-free sector and consider the Hamiltonian without $h_B$.
Therefore, we choose the initial state as $|\psi^{(0)}\rangle  = \otimes_i |+\rangle_i$, which is the only product state satisfying $ \langle h_B \rangle = +1$.
The GPEPO then only need to take the competition between $h_A$ and $h_x \sum_i \sigma^x_i$ into account.

To construct the GPEPO for the TCX model, we note that the local Hamiltonian consists of the four-site instead of the two-site interactions.
Similar to the $h_{ij}^{(2)}$, one can express the $\alpha$-th power of the four-site interactions $h_{ijkl}^{(4)}$ as the contractions of the four four-leg tensors:
\begin{equation}
\begin{split}
 (h^{(4)}_{i,j,k,l})^\alpha = &  (A^\alpha_{v_1 v_2})_i \otimes (B^\alpha_{v_2 v_4})_j \otimes (C^\alpha_{v_1 v_3})_k \otimes (D^\alpha_{v_3 v_4})_l,
 \end{split}
\end{equation}
where the Einstein summation convention is assumed for the contractions of virtual indices $v_i = 0,...,D_\alpha-1$ [Fig.~\ref{fig:toric_code}(b)].
Since $h_A$ is  {$C_{4v}$} symmetric, we choose $A^\alpha_{v v'} = B^\alpha_{v v'}  = C^\alpha_{v v'}  = D^\alpha_{v v'} $.
It is then straightforward to identify $n = 2$ and $D_\alpha = 1,\ \forall \alpha$ with $A^0_{00} = I, A^1_{00} = \sigma^z$. 
Letting $A_{v v'} = \oplus A^\alpha_{vv'}$, the GPEPO with bond dimension $D = 2$ can be constructed as 
\begin{equation}
G_{v_1 v_2 v_3 v_4} \sim \overline{A_{v_1 v_2} A_{v_3 v_4} }/2!
\end{equation}[see Fig.~\ref{fig:toric_code}{(c)}].
%
Similarly, the one-site interaction can be incorporated into the GPEPO as  
\begin{equation}
G_{v_1 v_2 v_3 v_4} \sim \sum_{\beta=0}^{1} \overline{A_{v_1 v_2} A_{v_3 v_4} C^\beta}/3!,
\end{equation}
with $C^0 = I,\ C^1 = \sigma^x$.
Enforcing the {$C_{4v}$} symmetry and the normalization condition, the non-zero action of GPEPO with $4$ parameters can be written as
\begin{equation}
\label{eqn:toric_GPEPO}
\begin{split}
G_{0000} &= I + c_3 \sigma^x,\  G_{1100} = G_{0011} = c_1\sigma^z\\
G_{1111} &= c_2I + c_4\sigma^x/3
\end{split}.
\end{equation}

\begin{figure}
\includegraphics[width=\linewidth]{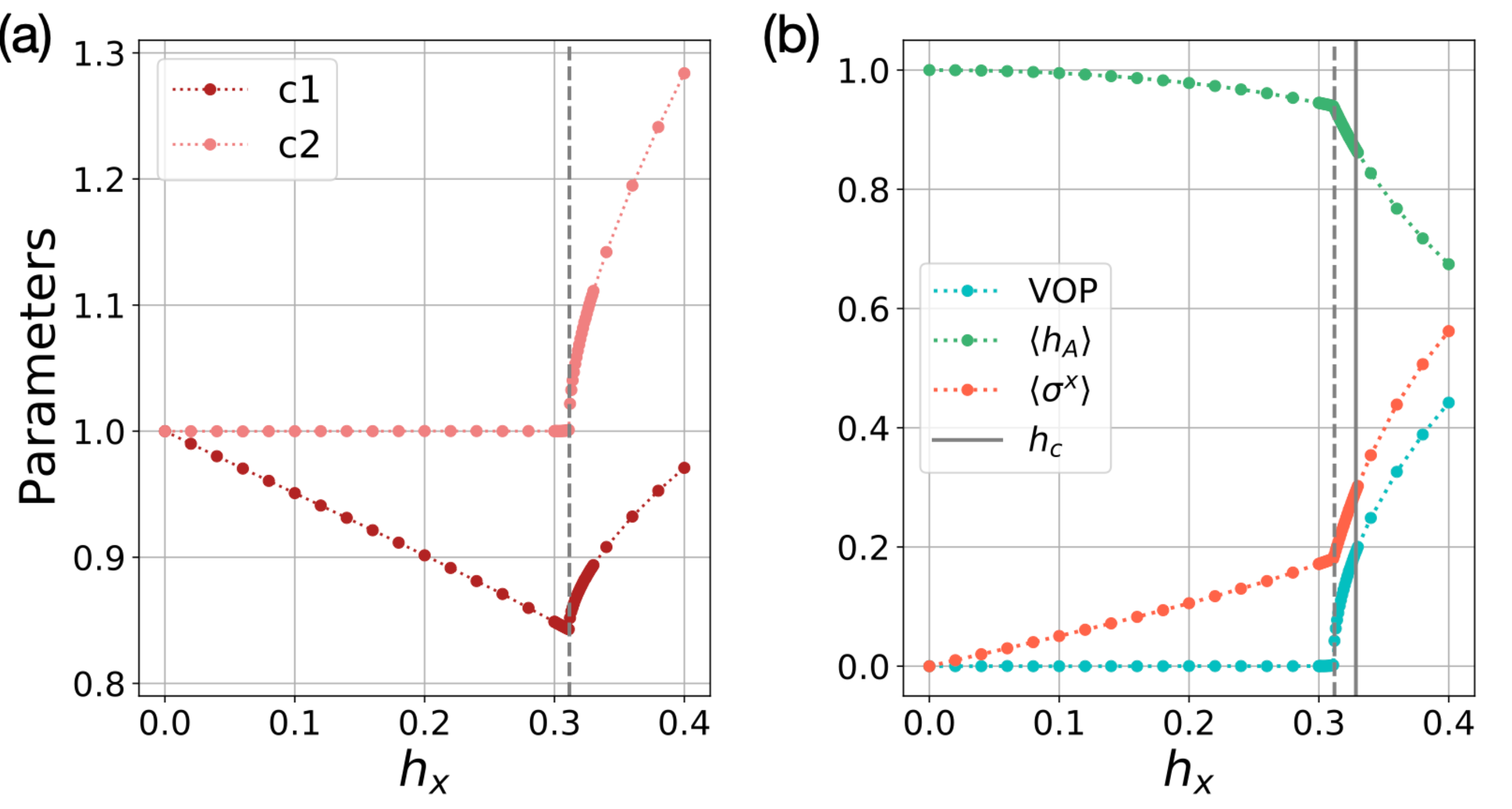}
\caption{GPEPO study of the TCX model. (a) The evolution of the variational parameters for $|\psi^{1} \rangle$ as a function of  $h_x$. (b) VOP, $\langle h_A \rangle$, and $\langle \sigma^x \rangle$ as a function of the magnetic field for $|\psi^{(1)}\rangle$.} 
\label{fig:toriccode_data}
\end{figure}

Fig.~\ref{fig:toriccode_data}(a) shows the evolution of the parameters as we increase the magnetic field for the first-order GPEPS $|\psi^{(1)}\rangle = \hat{G}|\psi_{0}\rangle$. 
Since $\sigma^x_i|\psi_{0}\rangle = |\psi_{0}\rangle$, $c_3$ and $c_4$ can be discarded.
The phase transition point at $h_x \approx 0.3115$ is clearly identified by the abrupt change of the parameters.
To identify the VOP, we note that at $h_x = 0$, the exact toric code ground state corresponds to the application of the GPEPO with $c_1 = c_2 = 1$. 
In this extreme limit, the GPEPO satisfies the extra virtual symmetry
\begin{equation}
\includegraphics[width=0.45\linewidth]{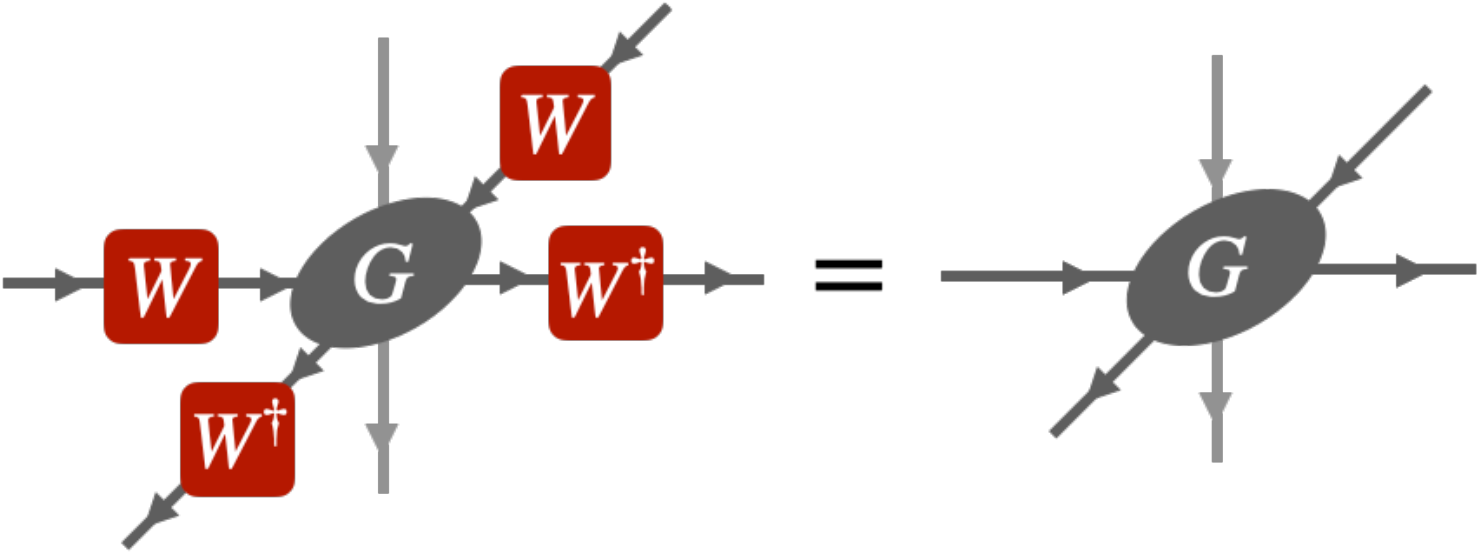},
\end{equation}
where $W = \sigma^x$. 
It turns out that this virtual symmetry is the manifestation of the $\mathbb{Z}_2$ TO phases \citep{Schuch_2010_Ginjective} [see App.~\ref{App:EOP} for details].
The VOP can then be identified as the difference between the symmetry-transformed and the original GPEPS.
Fig.~\ref{fig:toriccode_data}(b) shows the VOP as well as the physical observable $\langle h_A \rangle$ and $\langle \sigma^x \rangle$. 
One can clearly observe the happening of the phase transition at $h_x \approx 0.3115$. 
{
We note that given the PEPS tensor respecting the virtual $\mathbb{Z}_2$ symmetries, there are several ways to calculate the topological entanglement entropies \citep{Cirac_2011_ES,Orus_2014}, entanglement spectrums \citep{Schuch_2013}, and even extract the low-lying anyonic excitations \citep{Haegeman_2015,Chen_2022}. 
The optimized GPEPS can be directly served as an input tensor of those framework without any gauge transformation, which demonstrates the compatibility of GTNO with other numerical schemes. 
}

To get a more accurate ground state, we consider the second-order GPEPS with $4$ additional parameters.
The VOP and physical observables show similar behavior as $|\psi^{(1)}\rangle$, and the critical point is identified at $h_x \approx 0.3265$, which is already rather close to the result of QMC. 
Tab.~\ref{table:TC_result} shows the comparison of energies at $h_x = 0.2,\ 0.36$ and the estimated critical points obtained from GPEPS and other methods. 
For $D = 2$, with merely two parameters, GPEPS clearly outperforms the perturbation results ($1$ parameter) on locating the critical point. 
Furthermore, GPEPS yields comparable energy with the VarOpt method (roughly $32$ parameters), indicating that GPEPS reduces the number of parameters efficiently without losing the entanglement structure for the TCX model.
For $D = 4$, with  six parameters, GPEPS {has} identified a more accurate critical point than the perturbation method ($15$ parameters) as well. 
Besides, it {achieves} a comparable energy with the VarOpt ($\approx 54$ parameters), and even locates the critical point more accurately than VarOpt.
All of these comparisons strongly demonstrate the power of GPEPS.
We also evaluate the charge condensation order parameters introduced in Ref.~\citep{Iqbal_2021_orderparam}.
The slope of the scaling close to the critical point matches the critical exponent $\beta \approx 0.3265$ of the order parameters of the 3D Ising transition, consistent with results reported in Ref.~\citep{Iqbal_2021_orderparam}(see App.~\ref{App:EOP} for details).

\begin{table}
\caption{
The energies and the predicted critical points of TCX model compared to VarOpt \citep{Crone_2020_toriccode} and the perturbation method \citep{Vanderstraeten_2017_perturbation}. The critical point is located at $h_c = 1/3.04438(2) \approx 0.3285$ as estimated by the Monte Carlo simulations \citep{Blote_2002_MCTFIM} using the mapping between TFI and TCX models. 
}
\begin{tabular}{c c c c c}
 \hline  \hline
 & \text{num of params} & $h_x = 0.2$ & $h_x = 0.36$ & $ h_c$\\ 
 \hline 
GPEPS (D = 2)& 2 & -1.010266 &  -1.041798 & 0.3115 \\ 
VarOpt (D = 2) & 14* & -1.010410 & -1.042160 & 0.3351 \\ 
Perturbation (D = 2) & 1 & N/A & N/A  & 0.25 \\
\hline
GPEPS (D = 4) & 6 &  -1.010418 & -1.042219 & 0.3265\\ 
VarOpt (D = 3) &  54* & -1.010420 & -1.042200 & 0.3331\\ 
Perturbation (D = 4) & 15 & N/A  &N/A  & 0.322\\ 
 \hline \hline
\end{tabular}

\raggedright{*under the assumption that only the mirror symmetry is taken into account.}
\label{table:TC_result}
\end{table}

\section{Discussion and Outlook}
\label{sec:discussion}

We have introduced the Generic-TNO as a unifying scheme to study the general quantum many-body  systems. 
GTNO combines the idea of ITE that the ground state can be obtained by applying some operators generated from the Hamiltonian, and VarOpt by allowing free parameters to optimize the wave function. 
Applying GTNO to  simple initial states, the resulting GTNS can be used to obtain reliable ground states with far less parameters than the usual TNS. 
After the optimization, symmetries of the local tensor emerge without performing any gauge transformation, providing the probes to classify  quantum phases. 

{
We demonstrate the application of GTNO by studying several models in one and two spatial dimensions.
For the 1D Heisenberg model, we obtain a $SU(2)$ symmetric GMPS with extremely small energy deviation $\delta E \approx 1.31\times 10^{-5}$ and large correlation length $\xi \approx 205$ unit cells using only six parameters.
For the 1D TFI (TFC) model, we adopt the $\mathbb{Z}_2$ ($\mathbb{Z}_2 \times \mathbb{Z}_2$) symmetric GMPS to search for the ground states and detect the phase transitions using the emergent symmetries of the local tensor.
The power of GTNO is further manifested when considering the 2D systems.
Applying the $SU(2) (U(1))$ symmetric GPEPO with merely $5 (16)$ parameters to study the 2D Heisenberg model, we obtain comparable energy with the VarOpt methods using $110$ parameters for the same bond dimension $D = 4$. 
For the 2D TFI model, we allow the GPEPS favors one of the degenerate ground states in the SSB phase to compare with other numerical approaches. 
This shows the flexibility of GPEPO to combine with the initial states containing variational parameters.
While we currently have no general recipe to deal with the Hamiltonian consisting of four-site interactions, we explicitly construct the GPEPO for the TCX model.
Remarkably, the GPEPS with merely six parameters and bond dimension $D = 4$ already estimate more accurate critical points than the VarOpt with $54$ parameters for $D = 3$.
}


There are many possible extensions of the present work.
First of all, in this paper, we focus on the Hamiltonian consisting of the Pauli matrices, as they all guarantee the closed condition that $\{ (A^\alpha|B^\alpha ) \}$ with $\alpha = 0,1,...,n-1$ form a basis of any polynomial function of $h^{(2)}_{i,j}$. 
This property also holds for several models consisting of other operators, such as the Clifford algebra $\Gamma^\alpha$ satisfying $\{ \Gamma^\alpha, \Gamma^\beta \} = 2\delta_{\alpha \beta}$ {\citep{PhysRevLett.102.217202}} and the {clock and shift matrices defining the quantum $Z_n$ clock model}~\citep{Rajabpour_2007}. 
{As a result,} the construction of GTNO can be directly applied.
Furthermore, we note that the requirement of the closed condition is to guarantee that GTNO can obtain exact ground state for the frustration-free Hamiltonian.
For the model not fulfilling the closed condition, one can still use the power of $h^{(2)}_{i,j}$ to generate more physical operators, and we believe that the GTNO framework can still yield accurate ground state with far less parameters than the usual TNS. 

Second, GTNO can be used to study several exotic and numerically challenging models whose theoretical PEPS frameworks have been developed.
For example, the general string-net and 2D SPT models have been classified using the matrix-product operator (MPO) symmetries in the PEPS framework \citep{Williamson_2016_SPT,Bultinck_2017_anyonMPO}. 
However, variationally finding the optimal ground states with desired gauge symmetries is still challenging due to the multi-site interaction of the local Hamiltonian, which limits the current numerical study to the local-filtering method \citep{Francuz_2020_determineNAB,Marien_2017}. 
By generalizing the GTNO to these models, one should be able to accurately study their topological phase transitions.

Third, by taking  advantage of the fact that the symmetries of the local tensor {automatically} emerge during the optimization, GTNO can be used to explore the models whose PEPS frameworks are not {yet} well understood.
For instance, whether PEPS can faithfully represent the chiral shin liquids (CSL) is still under heated debate.
While several efforts are devoted to representing gapped CSL using PEPS, they all exhibit infinite correlation length \citep{Wahl_2013,Wahl_2014,Poilblanc_2015,Poilblanc_2016,Poilblanc_2017,Chen_2018,Chen_2020,Lee_2020,Chen_2021}, consistent with the no-go theorem proven in the free-fermion case that the parent Hamiltonian of a chiral PEPS is gapless \citep{Dubail_2015}.
Nonetheless, in a very recent paper, Hasik \textsl{et al.}  argue that a faithful representation of the CSL phase is in fact possible \citep{Hasik_2022}.
It is interesting to apply our GTNO framework to study the similar models in Ref.~\citep{Hasik_2022} and see whether there will {be} some extra symmetries emerging in the CSL phase.

Finally, we note that GTNO can also be used to study the Kitaev's honeycomb model \citep{kitaev_2006}, and one can show that the construction is similar to the dimer gas operator proposed in Ref.~\citep{Lee_2019}.
However, the distinguishing feature of the Kitaev spin liquids is that there is a constant of motion called vortex associated with each plaquette.
The current GTNO cannot be used to project the wave functions to the vortex-free sector, and one needs the loop gas operator introduced in Ref.~\citep{Lee_2019}. 
Figuring out how to systematically construct the vortex-free projector of the several generalized Kitaev models \cite{Barkeshli_2015,Chulliparambil_2020_16fold} and applying GTNO to probe the accurate ground state properties are currently under investigation.

\acknowledgments

Y.-H.C., K.H and Y.-J.K. are supported by the Ministry of Science and Technology (MOST) of Taiwan under grants No. 108-2112-M-002-020-MY3, 110-2112-M-002-034-MY3, 111-2119-M-007-009, and by the National Taiwan University under Grant No. NTU-CC-111L894601. W.-L.T. and H.-Y.L. are supported by National Research Foundation of Korea under the grant numbers NRF-2020R1I1A3074769. H.-Y.L. was supported by Basic Science Research Program through the National Research Foundation of Korea(NRF) funded by the Ministry of Education(2014R1A6A1030732). W.-L.T. thanks the additional computational resources provided by the Institute for Solid State Physics~(ISSP), the University of Tokyo.

\appendix
\section{Entanglement Order Parameters}
\label{App:EOP}
In Sec.\ref{subsec:1D_TFIM} and Sec.\ref{subsec:TCXM}, we find that the virtual $\mathbb{Z}_2$ symmetries of the local GMPS and GPEPS naturally emerge in the 1D SSB and 2D TO phases, respectively.
The virtual order parameters (VOP) signaling the phase transitions are then identified as the norm difference between the original and symmetry-transformed tensors.
Remarkably, we can defined another quantities, called \textit{entanglement order parameters} (EOPs), to detect the phase transitions and further study their critical behaviors.
The idea of EOP is originally introduced in Ref. \citep{Haegeman_2015} as a probe to model anyons' behavior by physically deforming the toric code wave functions to the trivial phases.
Since then, it has been generalized to the Abelian double models \citep{Duivenvoorden_2017}, string-net models \citep{Marien_2017}, and even been used to identify quantum spin liquids nature of the spin-$1$ Kitaev model \citep{Chen_2022}.
Very recently, Iqbal \textsl{et al.}\ have further shown that one can use the EOP to extract the critical exponents near the phase transitions \citep{Iqbal_2021_orderparam}.
In the following we give an intuitive explanation on why the tensors satisfying this property, called $\mathbb{Z}_2$-invariant MPS and PEPS, form a natural framework to represent the 1D $\mathbb{Z}_2$ SSB and 2D TO phases \citep{Schuch_2010_Ginjective}. 
After that, we motivate the construction of the {{EOP}} \cite{Iqbal_2021_orderparam} and show the results using the tensor optimized by GMPO and GPEPO.
We refer interested readers to Refs.~\cite{Schuch_2010_Ginjective,Iqbal_2021_orderparam} for details.

\subsection{$\mathbb{Z}_2$-invariant MPS}
\begin{figure}
\includegraphics[width=\linewidth]{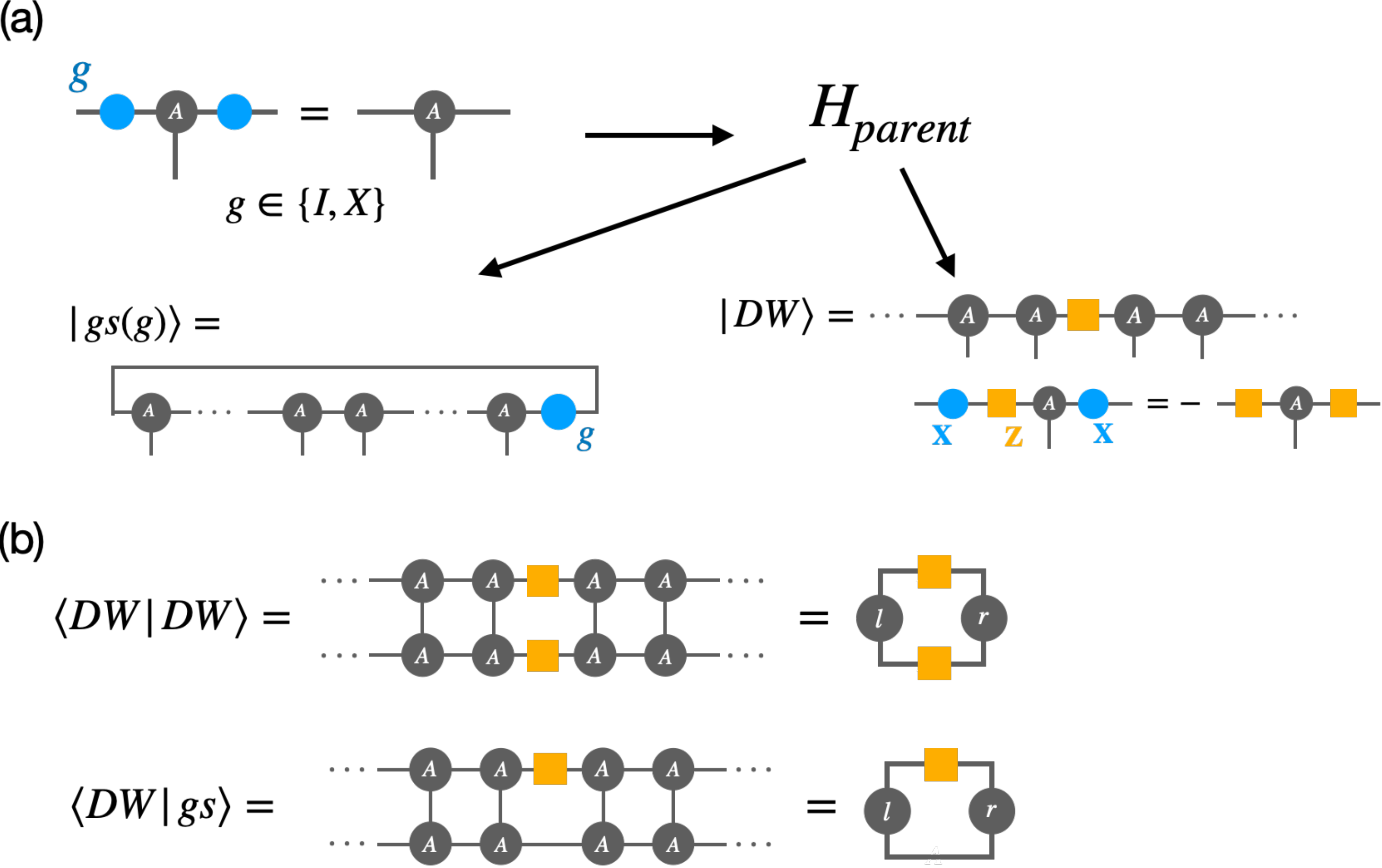}
\caption{
(a) The framework of $\mathbb{Z}_2$-invariant MPS. 
(b) The normalization of the domain-wall ansatz and the overlap between the domain wall and the ground state can be transformed into the local virtual observables, i.e., the entanglement order parameters.
Here, $l$ and $r$ denotes the effective environment of the MPS.
} 
\label{fig:Z2_MPS}
\end{figure}

\begin{figure}
\includegraphics[width=\linewidth]{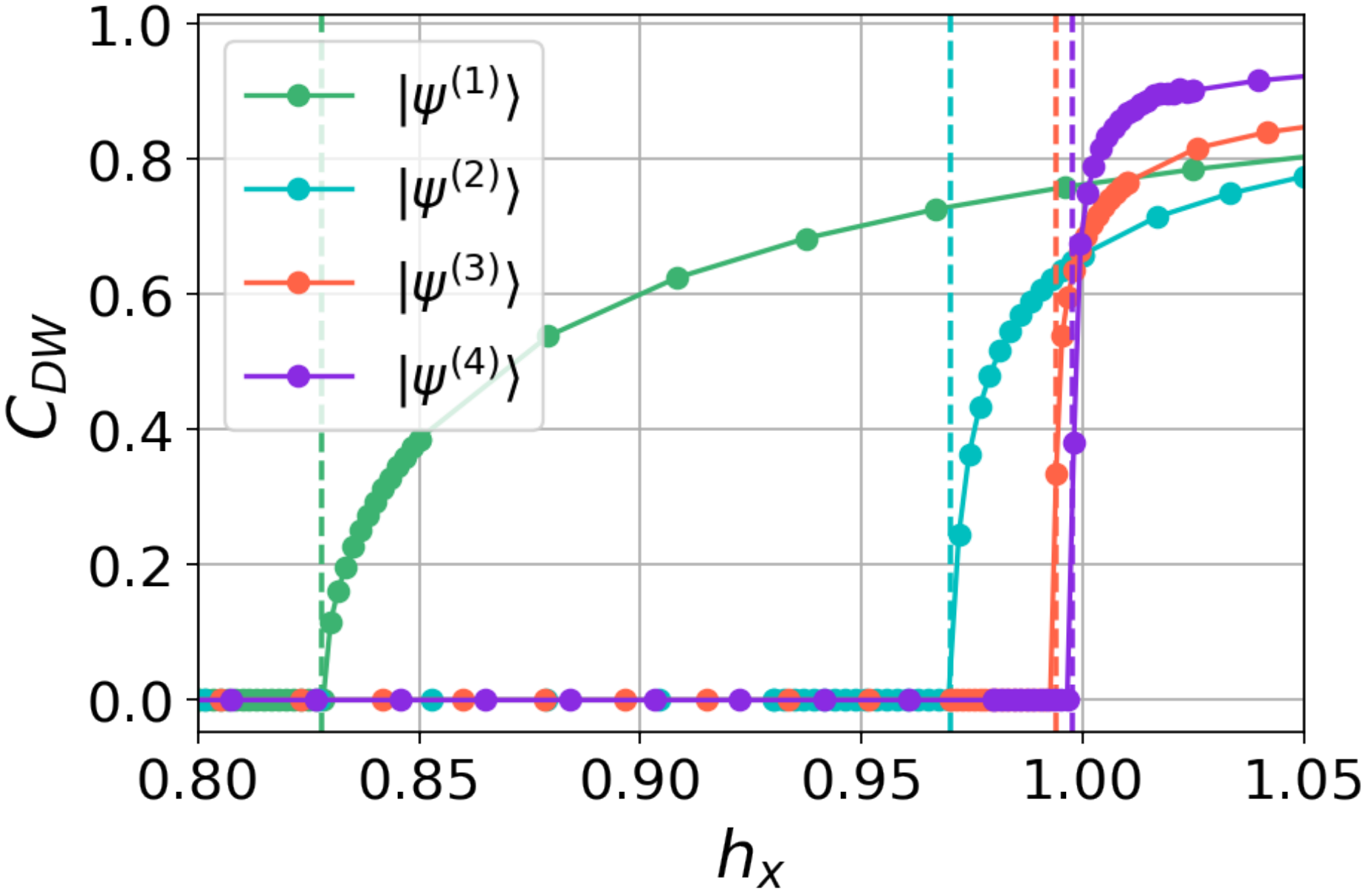}
\caption{The domain wall condensation order parameter as a function of magnetic field for $|\psi^{(s)}\rangle,\ s = 1,...,4$.} 
\label{fig:TFI_XI}
\end{figure}

Given a $\mathbb{Z}_2$ invariant MPS satisfying $u_g A u_g = A$ with $g = \{I,X\}$ and $u_g$ the representation of $g$, one can construct the parent Hamiltonian such that its ground state subspace can be obtained by attaching a single $u_g$ on the virtual bond.
To simplify the notation, we denote $u_I(u_X)$ as $I(X)$ and the operator that anti-commutes with $X$ as $Z$ in the following.
The degeneracy stems from the fact that the parent Hamiltonian cannot distinguish those states locally, as one can use the $\mathbb{Z}_2$-invariant property to {translate} the position of $u_g$.
Besides, we can create a domain wall excitation on top of the ground state by attaching a single $Z$ operator on the virtual bond of the $\mathbb{Z}_2$-invariant MPS.
Since the resulting tensor transforms anti-invariantly instead of invariantly under the virtual $\mathbb{Z}_2$ action, one can show that no local physical operator can create a single domain wall, which is a feature of the topologically non-trivial excitation.
We summarize the framework of $\mathbb{Z}_2$-invariant MPS in Fig.~\ref{fig:Z2_MPS}(a).
For the $s$-th order GMPS studied in Sec.\ref{subsec:1D_TFIM}, $X^{(s)} = I^{\otimes (s-1)} \otimes \sigma^x$ and $Z^{(s)}=I^{\otimes (s-1)} \otimes \sigma^z$  in the $\mathbb{Z}_2$ SSB phase.

However, $\mathbb{Z}_2$-invariant MPS does not necessarily guarantee the $\mathbb{Z}_2$ symmetry breaking phase, as the domain wall ansatz we create can either be confined ($\langle DW |DW \rangle =0$) or condensed ($\langle DW |gs \rangle =0$).
Therefore, one should further examine its normalization and overlap with the ground state.
In other words, the system is in the $\mathbb{Z}_2$ SSB phase only when the \textit{domain wall confinement order parameter} $K_{\text{DW}} = \langle DW |DW \rangle \neq 0$ and the \textit{domain wall condensation order parameter} $C_{\text{DW}} = \langle DW |gs \rangle =0$.
Using the fact that the single $Z$ action can be used to create a domain wall, $K_{\text{DW}}$ and $C_{\text{DW}}$ can be transformed into the virtual observables as shown in Fig.~\ref{fig:Z2_MPS}(b). 
One then use those observables to detect the phase transitions and further study the critical behavior.

Now, we use the GMPS optimized by GMPO in Sec.\ref{subsec:1D_TFIM} to calculate the EOPs.
Throughout the calculations, we find that $K_{\text{DW}} $ remains nonzero for all $h_x$. 
This is consistent with the fact that applying the transverse field will induce the phase transition only through the condensation.
Fig.~\ref{fig:TFI_XI} shows the domain wall condensation order parameter $C_{\text{DW}}$ as a function of transverse field for the $s$-th order GMPS with $s = 1,...,4$.
One can clearly observe that $C_{\text{DW}}$ signals the occurrence of the phase transitions, and the transition points coincide with the one identified using the VOP in the main text.
Besides, instead of getting flatter like the VOP, $C_{\text{DW}}$ becomes sharper as the bond dimension increases, showing its advantage of studying critical behaviors over the VOP.
We note that the GMPS does not enjoy the virtual $\mathbb{Z}_2$ symmetry in the symmetry-preserved phase, and the physical interpretation of $K_{\text{DW}}$ and $C_{\text{DW}}$ in this regime is unclear.
However, it is remarkable that the framework developed under the strict assumption of the virtual $\mathbb{Z}_2$ symmetry can still be applied to GMPO. 
Besides, throughout the calculations, we do not perform any gauge fixing condition as implemented in Ref.~\citep{Iqbal_2021_orderparam}. 
All of these demonstrate the power and flexibility of our GMPO framework.

\subsection{$\mathbb{Z}_2$-invariant PEPS}
\begin{figure}
\includegraphics[width=\linewidth]{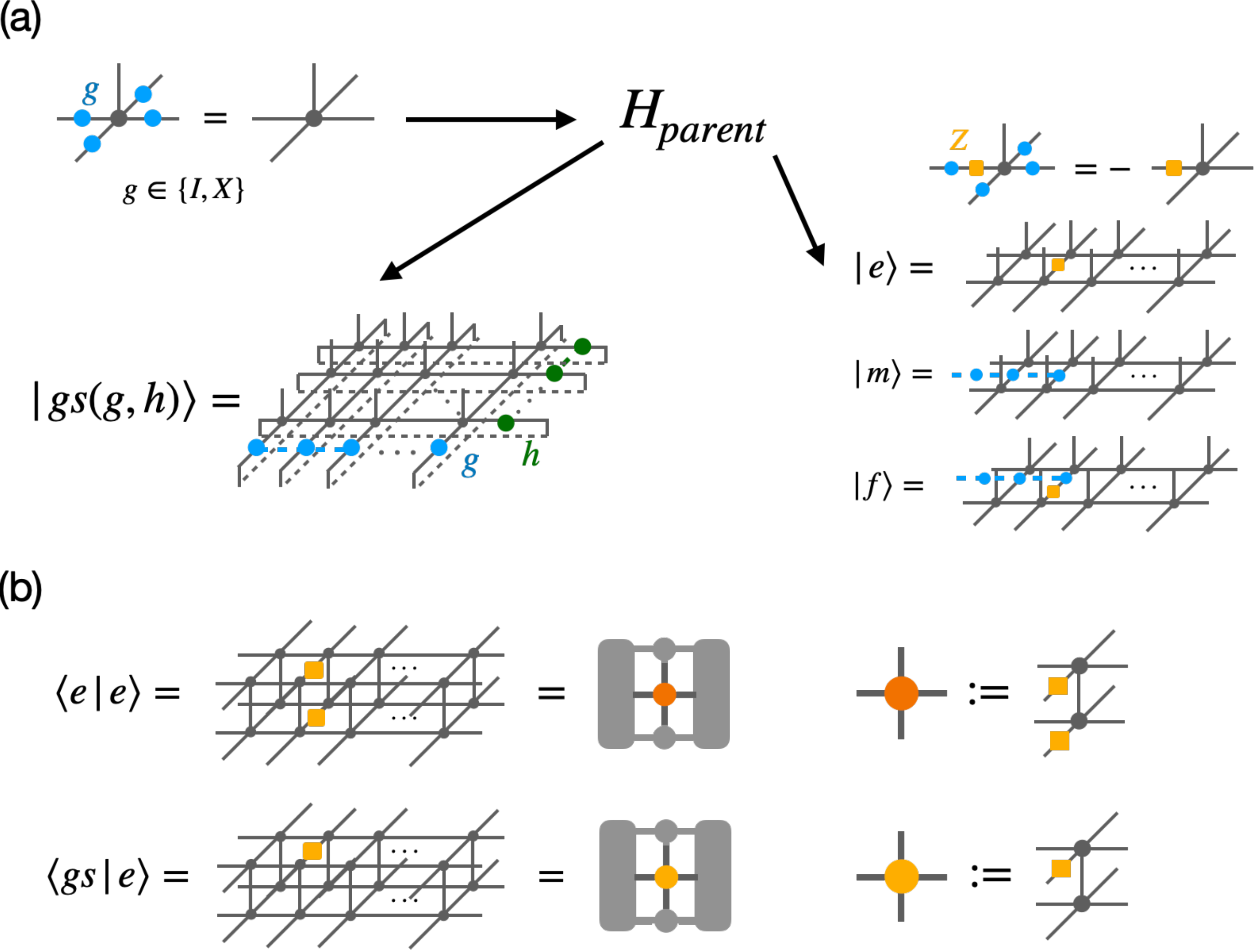}
\caption{
(a) The framework of $\mathbb{Z}_2$-invariant PEPS. 
(b) The normalization of the charge anyon ansatz and the overlap between the charge anyon and the ground state can be transformed into the local virtual observables, i.e., the entanglement order parameters.
Here, the grey patches denote the effective environment of the PEPS.
} 
\label{fig:Z2_PEPS}
\end{figure}

\begin{figure}
\includegraphics[width=\linewidth]{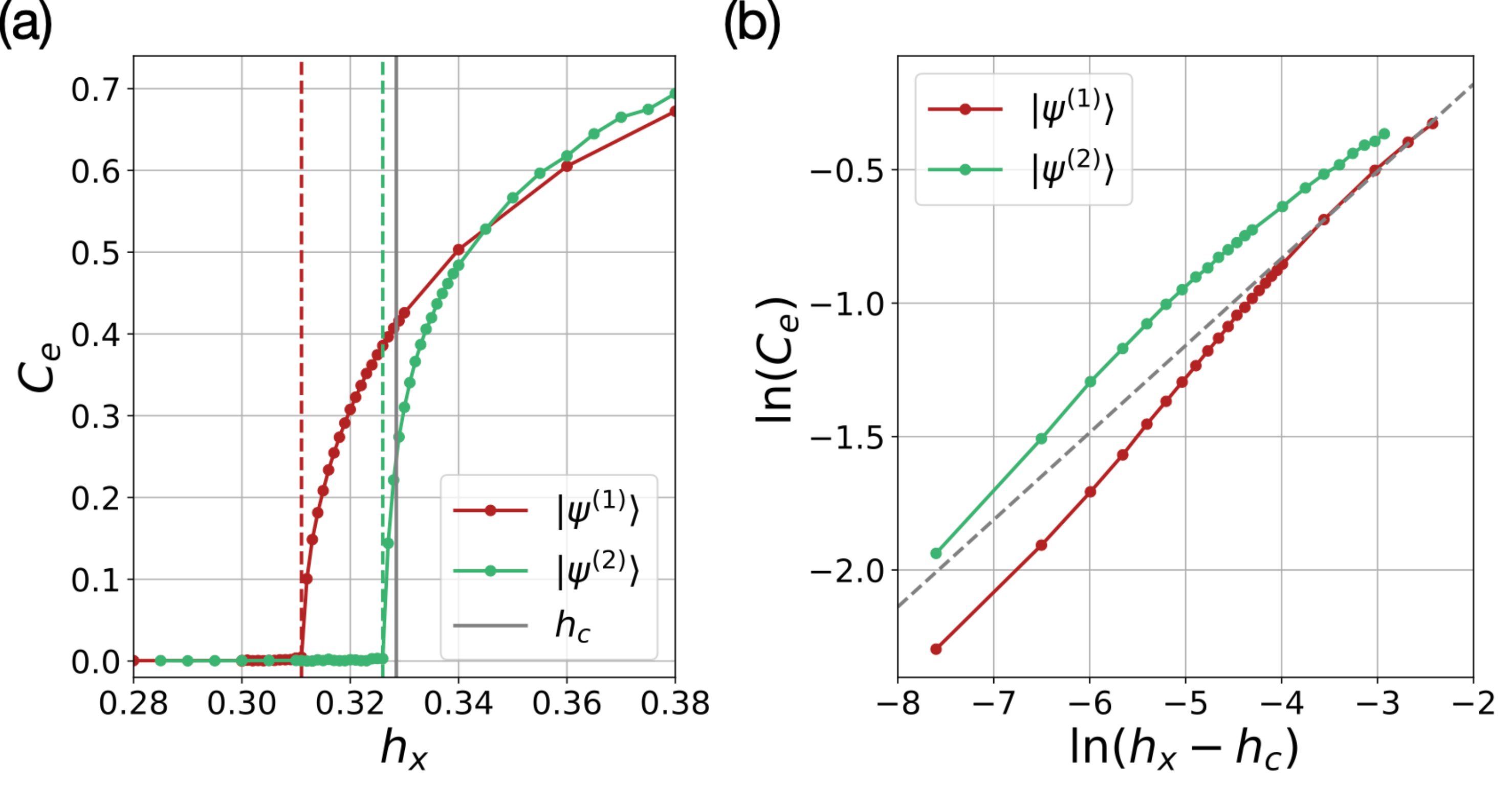}
\caption{(a) The charge condensation order parameter as a function of magnetic field. (b) Scaling of the charge condensation in the vicinity of the critical points. The slope matches the critical exponent $\beta \approx 0.3265$ of the order parameters of the 3D Ising transition.} 
\label{fig:TC_XI}
\end{figure}

{Parallel to the one dimensional counterpart, given} a $\mathbb{Z}_2$-invariant PEPS satisfying $A(u_g \otimes u_g \otimes u_g \otimes u_g) = A $ with $g = \{I,X\}$ and $u_g$ the representation of $g$, one can construct a parent Hamiltonian such that its ground state subspace is spanned by two non-contractible loop operators on the torus.
{Similar to} the $\mathbb{Z}_2$-invariant MPS, the degeneracy can be understood by noting that one can use the $\mathbb{Z}_2$-invariant property to {translate} the loop operators.
Furthermore, we can create anyonic excitations of the parent Hamiltonian on top of any four ground states.
The charge anyon $|e\rangle$ can be created by attaching a single $X$ operator, and the flux anyon $|m\rangle$ can be created by a half-infinite virtual $Z$ string on the virtual bonds of the $\mathbb{Z}_2$-invariant PEPS.
Finally, the fermion $|\epsilon\rangle$ can be created using the braiding rule $\epsilon = e\times m$ of the $\mathbb{Z}_2$ QSLs.
We summarize the framework of $\mathbb{Z}_2$-invariant PEPS in Fig.~\ref{fig:Z2_PEPS}(a).
For the $s$-th order GPEPS studied in Sec.\ref{subsec:TCXM}, $X^{(s)} = I^{\otimes (s-1)} \otimes \sigma^x$ and $Z^{(s)}=I^{\otimes (s-1)} \otimes \sigma^z$ in the $\mathbb{Z}_2$ TO phase.
Similar to the $\mathbb{Z}_2$-invariant MPS, $\mathbb{Z}_2$-invariant PEPS does not promise the $\mathbb{Z}_2$ TO phase, and one should investigate the normalization of the different anyonic excitations and their overlaps with the ground state.
Remarkably, since the condensation of the charge(flux) anyon is always accompanied by the confinement of the flux(charge) anyon, the TO phase can be identified if the charge confinement order parameter $K_e = \langle e|e \rangle$ is nonzero and the charge condensation order parameter $C_e = \langle e|gs \rangle$ vanishes.

Now, we use the GPEPS optimized by GPEPO in Sec.\ref{subsec:TCXM} to calculate the EOPs.
Throughout the calculations, we find that $K_{e} $ remains nonzero for all $h_x$. 
This is consistent with the fact that applying large enough $h_x$ will induce the phase transition through the charge condensation.
Fig.~\ref{fig:TC_XI}(a) shows the charge condensation order parameter $C_{e}$ as a function of magnetic field for the 1st and 2nd order GPEPS.
As expected, non-vanishing $C_{e}$ indicates the happening of the phase transition, and the estimated critical fields are consistent with the ones using VOP in Sec.\ref{subsec:TCXM}.
To demonstrate that GPEPS can also be used to study the critical behavior even if it does not respect the virtual $\mathbb{Z}_2$ symmetry in the charge condensed phase, we perform the log-log scale of $C_e$ and $h_x$ in Fig.~\ref{fig:TC_XI}(b). 
We find that both the slopes of 1st and 2nd order GPEPS match the the critical exponent $\beta \approx 0.3265$ of the order parameters of the 3D Ising transition.
This is consistent with the results reported in Ref.~\citep{Iqbal_2021_orderparam} even if we do not strictly impose the $\mathbb{Z}_2$ symmetry and have extremely few free parameters (two and six parameters for the 1st and 2nd order GPEPS, respectively).

\bibliography{bibs}

\end{document}